\documentclass[11pt]{article}   	
\usepackage{setspace}
\usepackage{geometry}                		
\usepackage{amsfonts}
\usepackage{amsmath}
\usepackage[parfill]{parskip}    		
\usepackage{graphicx}				
\usepackage{amssymb}
\usepackage{tikz}
\usetikzlibrary{shapes, arrows}
\tikzstyle{process} = [rectangle, draw, text centered, minimum height=2em]
\tikzstyle{connector} = [draw, -latex']

\title{Random trajectory models for complex phenomena}
\author{Jeffrey D Picka}
\date{}							

\begin{document}
\maketitle

\begin{abstract}

Many models for complex phenomena use a model for strongly-interacting elements on a small scale to generate larger-scale simulations of some aspects of experimental realizations. These models may be agent-based (as in the case of discrete element method models for granular flow) or based on pattern-forming systems of PDEs (as in models for Raleigh-B\`enard convection patterns). Often these models are purely deterministic, producing a single simulation for each set of initial conditions. If observed realizations demonstrate between-realization variability for important aspects of the phenomenon, those aspects can be simulated by adding probabilistic components to the deterministic models to create random trajectory (RT) models. 

The RT model framework provides probabilistic models which can be fit to data and validated, together with a clear perspective on how difficult it can be to establish any kind of validity for a fitted model. It treats models as code with adjustable coefficients, rather than as systems of differential equations. It provides a simply stated necessary condition for these code models to be easily fit and verified, as well as an argument that this condition can almost never be checked. When the necessary condition cannot be checked, the RT model framework identifies the code models as black box models which may have the capacity for emulating the joint distributions of small collections of statistics observed on realizations, but which can only provide very weak evidence for any form of explanation for the emergence of any aspect of the phenomenon. The framework also provides a way to clearly understand why finding a useful code model and scientifically validating it may require many person-years of extra experimentation and statistical analysis undertaken after the first output-producing code model is constructed and contributed. 

\end{abstract}

\section*{Introduction}

The validation of computer models for simulating complex phenomena requires confronting the many weaknesses that these models can have. In cases where the model simulates individual realizations of the complex phenomenon through the simple interactions of many smaller-scale elements, the random trajectory (RT) approach to modelling can be used to identify steps in the model-building process which require validation by data before explanatory claims can be made using model output. The random trajectory approach also demonstrates clearly where the hardest parts of validation lie, and it suggests what the costs of avoiding these steps can be. 

One of the simplest complex phenomena that can be represented by random trajectory modelling is a granular flow \cite{andfort}. When a powder flows, it behaves like a fluid in some ways but it is not amenable to continuum modelling. Instead, it is necessary to simulate the flow by simulating the motion of each individual grain in the flow as well as the interactions of the grains with each other, with the fluid which surrounds the grains, and with any confining surfaces. The construction of the code for the flow simulation requires ignoring or simplifying much of what is known about grain interactions in order for the code to be able to produce simulations within a reasonable amount of time on available hardware. Careful consideration of how a simulation model is constructed will show how difficult it is to fit values for parameters based on emergent properties of the flow, and why any such fitted model needs to demonstrate its predictive or explanatory capacity on multiple samples of many new experimental realizations. 

RT modelling can be applied to any phenomenon which is simulated by a solver for a system of differential equations that  is sensitive to initial or boundary conditions, or any agent-based model with similar sensitivities. Solver-based models are necessary for the representation of all turbulent fluid flow phenomena \cite{zikanov} (including the weather, ocean currents, and planetary climates) and for the representation of pattern-forming phenomena represented by systems of partial differential equations \cite{hoyle} (such as vegetation patterns in arid regions, emulsions, crack-based tiling structures, and cells in Rayleigh-B{\'{e}}nard convection). Agent-based models are needed to represent granular flows, the structure and formation of composite materials (including concrete, fibre-reinforced composites, and non-woven fabrics), large networks, the flocking behaviour of animals, and the spread of diseases through populations. In almost every case, individual realizations defy any simple summarization and no two observed realizations have the same microscale structure.

\section*{Simulations of sphere packings by RT modelling}

RT modelling will be defined via a relatively simple example of granular flow modelling: filling a cylinder with spheres to form a monosized sphere packing. The spheres to be packed will be assumed to have the same radius, the same composition, and uniform density. They will also be made from a material which is hard (resistant to wear) and tough (resistant to deformation). The container will also be assumed to be perfectly cylindrical with a flat bottom and to also be made from a hard and tough material. The upper part of the cylinder will have a hard and tough funnel attached which ensures that all poured spheres fall into the cylinder. Before the pouring begins, the cylinder and funnel will contain only air. The initial state of the spheres may be a packing within another cylinder which is then poured into the receiving cylinder, or may be some simpler but more arbitrary arrangement of spheres above the funnel. Experimental creation of packings of this kind can be undertaken indefinitely by anyone who has a funnel, a tempered glass cylindrical container, a small number of new hard plastic balls, and a clear, well-defined procedure for the filling process.

The RT modelling of a monosized sphere packing may be of direct interest, or useful as a component in a simulation model of other complex systems. In the mathematical or statistical study of the disordered packing of spheres \cite{cs:1999,zong:1999,picka:2012}, a sufficiently validated model for a packing of physical spheres could be used to efficiently simulate the emergence of packed structures under a wider range of conditions than are experimentally possible, and to investigate aspects of their structure that are impossible to experimentally observe. In the simulation of composite materials such as concrete, packings can be a starting point for simulating disordered arrangements of idealized rocks within a cement matrix \cite{picka:2000b}. By shrinking packed spheres and slightly perturbing their locations, it is possible to simulate patterns of inclusion that cannot be simulated by classical probabilistic methods for simulating disordered arrangements of non-overlapping spheres. Sphere packings can also be used to represent the initial state of packed metal spheres in a cylindrical mould in a simulation of a sintering process \cite{coube,henrich:2007}, or an initial state of packed spheres of a drug in a simulation of the process of compaction to form a tablet \cite{russell:2022}. 

\subsection*{Explaining the physical process of packing formation}

What happens during the experimental process of packing formation can be described by using well-established explanatory narratives from physics and materials science which are based on physical laws \cite{zhu}. 

When poured, the spheres will be acted on by gravity and will accelerate down into the cylinder. The spheres will collide with each other and with the containing surfaces via interactions which conserve mass, linear and angular momentum, and energy. The spheres will displace the air in the container, which will be forced upwards out of the container in a turbulent flow. When collisions occur, the spheres in contact will deform to a very small degree. When contact is released, the spheres will not quite return to their original shape. While steel spheres are spherical and possibly closer than any other material object to an ideal sphere, they become less spherical with each succeeding contact. Contacts also will involve surface friction, which will raise the temperature of the spheres and transfer heat to the surrounding air. Friction may result in fretting, in which small fragments break off of the spheres. These fragments may fall, may be entrained in the flowing air, or may adhere to the grain surfaces, where they can interfere with contacts between spheres. When spheres contact the container, the container will also be subject to irreversible deformation, friction, and possibly fretting and the accumulation of debris on its surface. 

If the spheres are very small, there can be attractive forces between them due to intermolecular forces, and also due to Coulomb forces arising from electrical charge being redistributed during contacts. Such forces will also bind or repel small fragments arising from damage to the spheres. 

The air that is displaced by the motion of the spheres is forced into turbulent flow through space between the falling spheres. The space occupied by the fluid is always changing until the very last stages of the process. The air is heated by viscous friction arising from turbulence and by transfer of heat from spheres and confining surfaces that was generated by material friction. The flowing air also exchanges momentum with the spheres through surface contact, and exerts forces acting on surfaces which can act in all directions. 

As the spheres settle into a packing, the location of each sphere depends on the locations of hundreds of its neighbours \cite{hales} and on the container surfaces. At the end of the process, the spheres are still and their kinetic energy has been expended by the work involved in permanently deforming and grinding the spheres, by heating the spheres and the air, by the work of the spheres on moving the air, and by transfer into potential energy stored in reversible deformations. The experimental packing can be made denser by tapping it, vibrating it, or rodding it. Historically, most simple experimental procedures for this kind of process resulted in packings with volume fractions close to 0.63 and no local order. More recently developed procedures have enabled denser and more locally ordered packings to be produced experimentally \cite{torquato:2000}.

It is possible to experimentally study the motion of a single sphere through the air, and to study how frictional or deformational contacts between pairs of spheres occur \cite{cavarretti:2011}. These studies involve specialized experimental settings which are very different from the phenomenon of packing formation. Within the packing process, the interactions between spheres, between spheres and the container, between the spheres and the air, and within different parts of the airflow itself are mostly unobservable. The final packed arrangement may only be observable via some physical process of decomposition \cite{bernal:1960,bernal:1970} or by some form of remote sensing \cite{reimann:2017}. In both cases, information about exact locations of spheres is not observable without error, and the cumulative effects of deformation and damage will be impossible to observe until after the packing is taken apart.

\subsection*{Building a formal model from the explanatory narrative}

To model an experimental packing formation process of this kind, it is necessary to build a formal mathematical model based on the explanatory narrative. This formal model will be a system of differential equations. Ideally, the spheres and container should be modelled as deformable bodies which, when subject to excessive force, can fracture to produce small particles or permanently deform. The fluid surrounding the spheres should be modelled as a continuum using a model that can represent the turbulence induced by the moving spheres. All changes in mass due to damage should be represented, as should be all energy flows between and within the solids and the fluids. The model would also be constructed to ensure that mass, momentum, and energy are all conserved. The model would be specified through systems of partial differential equations, and the solutions to these formal models would be mass, energy, and momentum fields. While such a formal model is conceivable, it would have no formal solution and any use of it would require an approximate solution by numerical methods. The deformation or fracture of a single sphere in contact with a wall and the flow a fluid around a single sphere are both very challenging problems in continuum modelling. Building a continuum model for the interaction of thousands of grains, a container, and the fluid in motion around these spheres that is faithful to all aspects of the explanatory narrative exceeds the capacity of any numerical modelling software that exists now or in the near future. 

For a formal model for granular flow to be useful, it is required that the model have a formal solution and a numerical approximation to that solution. This can be achieved by using a formal model which omits some interactions and aspects of the theoretical narrative from the model (abstraction) and by using non-representative simple models for the interactions which are to be included (idealization). The most common approach to constructing a model of this kind \cite{guo:2015} is by using a Discrete Element Method (DEM) model \cite{cundall}. The discussion of how these models are built is based on the detailed and critical descriptions of the construction of DEM models found in the work of Zhu \textit{et al.} \cite{zhu:2008} and O'Sullivan \cite{osull:2011}. 

In many models for granular flow, useful formal models are constructed by avoiding the faithful modelling of any aspect of the system than requires the use of spatial partial derivatives. The flow of the fluid around the grains is not modelled explicitly, although some simple grain-fluid interactions may be represented. The deformation and damage of the grains and the container are not directly represented by continuum models for deformable bodies, but are modelled by very simple mechanisms. When a strategy of this kind is followed, what would be a continuum model expressed as a system of non-linear partial differential equations becomes a model expressed as a system of non-linear ordinary differential equations (ODEs), with the only derivatives being taken with respect to time. These and further idealizations make the approximation of solutions by numerical methods feasible.

The main idealizations required to produce an ODE-based model are based on a soft shell model for the spheres. Conceptually, this involves imagining each spherical grain to consist of a rigid spherical core covered by a soft deformable shell of uniform thickness. This model for a grain was found to be necessary when DEM models based on hard spheres were unable to produce packings as dense as those seen in experimental packings of spherical objects. The soft shell model is not a continuum model which represents the soft shell as a deformable body. Instead, the spherical grain is represented as a point mass which moves according to its own inertia until affected by forces acting at a distance or buoyancy forces, or until the sphere becomes sufficiently close to another sphere or a container wall (which is also considered to have a soft coating). When sphere centres become close enough that their soft regions intersect but their hard cores do not, then multiple simple mechanisms appear which idealize the interactions undergone by spheres in contact. 

When the soft shells of a pair of spheres come into contact, four distinct forms of interaction can be represented. The spheres as objects are in translational motion, and this results in both a compressive force interaction along the vector connecting the sphere centres and a frictional force interaction resulting from motion in the plane orthogonal to that vector. Both spheres are also considered to be rotating along their own axes of rotation, and have angular momentum interactions. All of these interactions are not modelled as interactions between deformable objects, but instead are modelled by very simple mechanisms based on idealized springs, dashpots and rollers. The springs can represent the compressive and tensile forces which occur during collisions, but models based on springs alone simulate granular flows which never settle down to a stable final state. The dashpots in the simple mechanisms represent the energy losses associated with damage and deformations in the experimental phenomenon. The models for simple mechanisms are chosen so they could plausibly represent interactions analogous to those between physical spheres while still being part of a formal model which can be solved numerically. These mechanisms are not chosen to faithfully represent the flow and transformation of energy within an experimental realization of granular flow. 

The idealized formal model is a system of ODEs which can potentially be applied to many granular flows, once the coefficients are given numerical values through some form of model fitting. The formal model is based on a faithful description of the heavily idealized narrative, in which point masses interact via simple mechanisms when they are close enough to each other. Ignoring the technical details involving contact with the container, the model requires only nine values to be known for every one of the $N$ spheres at any one time: three values to specify the position, three values to specify the velocity, two values to specify the direction of the axis of revolution, and one value to specify the angular velocity. The velocities of each grain are related to their positions by $3N$ ordinary differential equations of the general form $v=dx/dt$. If at any instant the grains are in mechanical equilibrium, a further $6N$ ordinary differential equations are specified by force balances of the form:

\begin{align*}
m_i \frac{d \mathbf{v}_i}{dt} &= \sum_j \mathbf{F}_{ij}^c +\mathbf{F}_i^{cyl} + \sum_k \mathbf{F}_{ik}^{nc} + \mathbf{F}_i^f +\mathbf{F}_i^g  \\
I_i \frac{d \mathbf{\omega}_i}{dt} &= \sum_j \textbf{M}_{ij} + \mathbf{M}_i^{cyl}  \\ 
\end{align*}

On the left, $m_i$ represents the mass and $I_i$ the moment of inertia of sphere $i$. For this example flow, these would be constants and the subscripts can be dropped. The motion of sphere $i$ is described by its velocity $\mathbf{v}_i$ and  its angular momentum $\mathbb{\omega}_i$. On the right, $\mathbf{F}_{ij}^c$ is the contact force exerted on sphere $i$ by sphere $j$, $\textbf{M}_{ij}$ is the contact torque exerted on sphere $i$ by sphere $j$, and $\mathbf{F}_i^{wall}$ and $\mathbf{M}_i^{wall}$ are the corresponding forces associated with contacts between sphere $i$ and the cylinder. In some flows, there can be non-contact forces $\mathbf{F}_{ik}^{nc}$ exerted on $i$ by sphere $k$. There will also be forces arising from interaction with the surrounding air $\mathbf{F}_i^f $, and the force of gravity $\mathbf{F}_i^g$. 

In this formulation, there is no term to account for changes in angular momentum due to interaction with the air. Interactions with the container are assumed to be local and it is assumed that rigidity of the containing surfaces prevents one any one sphere/surface interaction from affecting any other. 

Once the system of ODEs is specified, further abstraction and idealization is necessary. If some of the forces are very small relative to others, it may be useful to omit them from the model. For the non-omitted forces, choices need to be made regarding which theoretical, empirical, or simple mechanistic model needs to be used. Again, these choices may be driven less by the desire to faithfully represent the force than by the need to produce a model that is numerically solvable with available computing resources within a reasonable period of time.

The need to build a model that can produce output may motivate further idealizations. It is far easier to work with a model for the powder which flows in two dimensions rather than three, in which the monosize spheres are replaced by monosize discs. Such models are easy to build, but all experimental flows are of three-dimensional physical entities. While two-dimensional models can be referred to as rod or disc models, they are not representing physical discs or rods. Long rods of cylindrical cross section are subject to flexing under their own weight or when subject to other forces. Thin discs will tend to escape from the plane unless subject to some kind of physical constraint in the third dimension. Assuming that the third dimension of a physical system does not need to be modelled is a very strong idealization that risks making the model incapable of representing any granular flow.

A second strategy for simplifying the model is to greatly simplify the ways in which the spheres interact with containing surfaces. Challenges with modelling those interactions can be avoided if it is assumed that there are no containing boundaries at all through the use of periodic boundary conditions. If the flow of spheres through an inclined trough of rectangular cross section is being modelled, any sphere whose centre is closer to the vertical trough side than its radius would have the part of the sphere that lies outside the trough appear within the trough on the opposite vertical side. The effect of this is to model the trough as an inclined plane of infinite extent which is tiled by identical strips of flowing spheres. While this strategy can be justified in some models for fluid flow and in models from kinetic theory, it is much harder to justify for a granular flow that is small enough to be modelled. Since granular flows exhibit significant interactions with constraining boundaries, this form of idealization has the potential to make the idealized formal model into one that cannot represent a physical phenomenon. 

Once the system of ODEs is specified, it is also necessary to ask whether any solutions exist to it, and whether or not those solutions are unique. Existence of solutions to initial value problems involving ODEs is theoretically straightforward. It is based on converting the system into a system of $n$ first order equations of the form 
\begin{equation*}
\frac{dX_i}{dt}=f_i(x_1, \ldots x_n, t)
\end{equation*}
and then showing that the $f_i$ satisfy a Lipschitz condition \cite{deuborn}. These proofs are made difficult by both number of non-linear equations involved and the fact that the ODEs constantly change in form and content as any particular modelled flow evolves in time. If an idealized formal model has no solution but can serve as the basis of output-producing code-based numerical model, then the ill-posedness of the idealized formal model may not matter if it inspires the creation of a useful numerical model. Ill-posedness does rule out using output from such a code model as evidence for an explanation that inspired the original formal model.

If a unique solution exists to the system of ODEs that constitute the idealized formal model, then that solution is a differentiable path through a $9N-$ dimensional space. Every point on that path is a list of $9N$ values describing the state of the system at the corresponding time. A numerical solution to the system is a sequence of points in that $9N-$ dimensional space in which every point in the sequence approximates the the value of the formal solution at some fixed time reasonably accurately. Constructing that numerical solution involves two entangled stages: the development of a discretized version of the idealized formal model, and developing an algorithm for producing an approximate solution to the discretized model for some set of initial values.

Numerical approximations to formal solutions are found by replacing the system of ordinary differential equations with a system of ordinary difference equations. Each derivative of the form $\frac{dx}{dt}$ is a replaced by a difference $\frac{x_{t+\delta t}-x_t}{\delta t}$, where $\delta t$ is the difference between times at which consecutive approximations $x_t$ and $x_{t+ \delta t}$ are calculated. The system of ODEs is replaced by a system of non-linear equations describing relations between the approximations at a sequence of times. There are many different approaches to carrying out this discretization process \cite{rougier:2004,poesch:2005,krugem:2008}, each of which will produce a different set of ordinary difference equations to solve. Each approach will produce a different approximation to the formal solution of the system of ODEs. There is no \textit{a priori} guarantee that any of these methods will produce a usefully accurate approximation to the solution of the idealized formal model. The formal statement of the chosen system of difference equations can be identified as the \textit{discretized formal model}. Further arguments are needed to establish that the discretized formal model can produce useful approximations to the solution of the formal model. 

The connection between the discretized formal model and the solution that it seeks to approximate can sometimes be established mathematically for an identifiable class of idealized formal models \cite{osull:2011}. This is necessary since no explicit formal solutions exist to the formal models, and so there is no correct value to which any specific approximation can be compared. Experimental observations of flows can play no role in the evaluation of the validity of this approximation. 

For simple ODE systems, it is possible to show that some discrete formal models will produce approximation sequences that converge to the formal solution in some suitable metric as the time step size approaches 0. The most useful proofs of this involve Lipschitz continuity, and so it is impractical or impossible to establish formally that the convergence occurs for the discrete idealized model or that the original idealized formal model has a solution. 

For an approximation to a formal solution to be useful, it also needs to be numerically stable.  It is possible for an approximation to be theoretically convergent, but also that the approximation diverges from the formal solution for any feasible time step size and becomes meaningless after a few time steps. It is known that the force balance equations in the formal DEM model are stiff, and so instability will occur unless the time steps are small enough to prevent its occurrence. One of the main challenges in building DEM models is to find a discretization method that is stable when using time steps which are large enough that the model can be usable. 

In the original method of implementing a discretization, several choices were made so that a DEM model could run on hardware available 50 years ago. The discretization was explicit rather than implicit, which meant that the approximation at time $t+\delta t$ was found from equations which were based only on terms of approximations at the previous time $t$. The original procedure was to establish which spheres are in contact at $t$, numerical solve the force balance difference equations for the acceleration of each particle at time $t+\delta t$, and numerically solve the equations to yield the velocities and positions at time $t+\delta t$ based on those new accelerations. This scheme is much more efficient than attempting to simultaneously solve simultaneously for all the variables at the new time, but also is not representative of how an experimental flow changes over time. With increasing hardware capabilities, simultaneous solution has become possible. The choice of which approach to use is based on the need to build a usable model given available hardware capabilities. 

As with other aspects of verification, there is no way to prove that a time step is small enough to guarantee stability over the length of time that the simulation is to run. Formal approaches rely on assumptions about the discretized formal model which cannot be checked, or which further idealize the discrete model before the analysis takes place. In some applications of numerical methods, lack of stability can be obvious when code based on the model produces output that clearly fails to represent the phenomenon. This approach to detection has been shown to fail in a very simple quasi-static DEM model for which a formal solution exists \cite{osull:2003}. Evidence of instability can be found by using the total system energy as a diagnostic statistic. If the total system energy increases over time in a way that can't be accounted for via modelled energy inputs, this is considered to be evidence of instability. 

Experimental investigations of instability can be undertaken using simulation models representing small numbers of monosized spheres \cite{otsubo:2017,burns:2018}. These studies suggest that the minimum step size needed depends inversely on the number of interacting spheres being modelled and inversely on the number of contacts between spheres. The presence of damping and the use of more realistic non-linear spring models in contact interactions also require reductions in the step size. As the step size needed to attain stability decreases, there is a risk of increased approximation error over time on account of needing to use more time steps. 

While it is conventional to treat the formal model as the representation of the phenomenon, that model cannot yield a formal solution of any kind which can be compared to experimental observations. If a discretized formal model cannot be verified, then it may not be capable of approximating the solution to the formal model. The discretized formal model also has no simply stated or easily constructed solution, and so it necessary to use it as the starting point for the development of code models which can be fit to data and which can produce output that can be compared to observations. 

\subsection*{From Discrete Formal Model to Code Model}

Since the discrete idealized model is a very large system of non-linear equations, its solutions also need to be approximated via calculations that are subject to round-off error. There are well-established numerical methods for doing these approximations, which are thought to introduce less error into the computed approximations than is introduced by problems with the stability and accuracy of the discretized formal model. The code also needs to include algorithms for adjusting time step size based on detection of known configuration-based issues and evidence of instability. These extra elements make the code into a model itself, distinct from the discretized formal model. A \textit{general code model} is code with adjustable coefficients which are inherited from the formal model. A general code model will be useful if it can rapidly produce stable output for some collection of coefficient values and some collection of initial conditions.

Practical use of the general code model also requires some method of specification of the initial conditions and of the coefficients inherited from the formal model. A \textit{specified code model} will be a general code model with numerical values for those constants set, which can produce a sequence of points in the $9N$-dimensional state space which describe the state of a simulated powder when given a suitable set of initial conditions. A specified code model is the first entity in this sequence of models which can produce anything that could potentially be compared to experimental observations. A \textit{fitted code model} will be a specified code model whose coefficient values have been chosen in order to make the model output consistent with aspects of observed flows.

If a fitted code model can produce output, it is potentially useful. Its utility cannot be inferred from its construction alone, since the idealizations, abstractions, and approximations that are need to produce the general code model may make it impossible for any specified code model to faithfully represent the physical process of packing formation. The utility of a fitted model needs to be established by some form of validation argument which involves a comparison between experimental observations and observations simulated from a specified code model. Validation can result in three possible outcomes. It could be that no specified model can be found which can produce output that in any useful way resembles observations of the phenomenon. If that is not the case, then the fitted model may be able to predict some aspects of what will be observed in new replicates of the experiment that produced the data used to fit it, in spite of misrepresenting the dynamics of the spheres in a physical powder flow in important ways. Such fitted models may prove to be useful for predicting the results of experiments not yet carried out. Any such predictive models would be useful in applied settings, but would be much less useful than models which can faithfully represent important aspects of the interactions of the physical grains in spite of all the idealization. If it could be clearly shown that a fitted code model could faithfully and reliably represent the grain interactions which cause important aspects of the granular flow, then the fitted code model could be considered explanatory. Any validation process for these models must be able to clearly distinguish these possible outcomes.

\subsubsection*{Augmenting the Code Model}

A fitted DEM code model is capable of producing exactly one simulated realization from a single set of initial conditions. To provide the capacity for a code model to represent experimental observations, it is necessary to augment the code by adding probabilistic elements. Two distinct augmentations are possible, and at least one of them must always be present in any model that requires validation using experimental observations.

If an experimental procedure for making a sphere packing is followed, no two packings (or the details of their formation process) will be the same for their individual grains. DEM models for such flows belong to a class of ODE models which have been found to be sensitive to initial conditions \cite{lorenz:1963} in the sense that any small change to the initial conditions produces unpredictable changes to the solutions. The behaviour of the model is used to explain experimental unpredictability by inferring that it arises from the inability of experimenters to control the exact initial states of the grains. When sensitivity to initial conditions is present, experimental realizations of a packing process will demonstrate between-realization variability in both details of the flowing grains and in some emergent larger-scale properties of the flow. This kind of variability cannot be captured by a deterministic model unless the unpredictability is represented by modelling the sensitivity to initial conditions.

If the sensitivity to initial conditions is not modelled, then it will be unclear as to exactly what a fitted DEM code model output can represent. It will not be possible to exactly identify the initial conditions for any one experimental trial, and so the fitted DEM code model cannot represent a single observed experimental realization. If it is claimed that the fitted DEM code model represents something that could be an experimental outcome, it will also need to be assumed that the initial conditions used were typical of what could happen in an experimental setting. If the chosen initial conditions are somehow atypical, then the deterministic DEM code model of the flow could represent a very unusual flow that might happen. The only way to avoid problems of this kind (other than assuming that they don't exist) is to augment the DEM model so that it can provide a sample of simulated flows which can be compared to a sample of experimentally realized flows. 

To build a model that can be validated, the DEM code model must be able to represent sensitivity to initial conditions. It is possible to add code that simulates repeating experiments with different initial conditions by randomly sampling different sets of initial states for the spheres. For this example, the spheres are assumed to be packed in a cylinder before the DEM model simulates the pouring. It would be necessary to develop a code model which can sample from an ensemble of packed arrangements of spheres which is similar to the collection of initial physical packings that could be made according the experimental procedure.

Constructing a cylinder packed with spheres in a lab can be seen as sampling in some way from an ensemble of cylinders packed with spheres that could be made via an experimental procedure by the particular group of researchers in that particular lab. A sample of initial states for an augmented code model could be generated by preparing packed cylinders and then using some form of remote sensing to estimate the locations of the individual spheres in those packings. Assuming that the remote sensing method has sufficient accuracy, this method would come closest to matching the behaviour of the experimental process.

To save on resources, it may be necessary to simulate the formation of the initial packed state. This is challenging, since the DEM model under construction is intended to represent that filling process. The simulated packings will not be from the same ensemble as observations of experimental packings. For a simulated initial packing to be useful, it is necessary to assume that differences between experimental and simulated initial states are not so different that it has an impact on the utility of the model that is being constructed. The validity of this assumption could be investigated by comparing code model output based on simulated initial states with code model output based on observations of experimental packings.

It may also be possible to save computing time and resources by using a packing-generating algorithm which has no connection to the physical process of packing formation. It is possible to assume that there is a well-defined ensemble of physical packings (designated as random close packings) which can be produced by a wide range of experimental procedures and which are characterized by having volume fractions of around 0.63 for the monosized spheres \cite{bernal:1959,finn:1970,finn:1970b}. If such an ensemble exists, then any algorithm capable of packing spheres at that volume fraction could be assumed to sample from it. Many algorithms of this kind were developed before DEM packing models became available \cite{vold:1964,vb:1972,jt:1985,barg:1991,lochmann}, but experimental studies of packing processes suggest that the ensemble of random close packings may not be well-defined \cite{torquato:2000,pouliquen,grebenkov}.

While adding a random sample of initial conditions is necessary to build a model that can be validated, there is a second way in which probabilistic modelling can be used to improve a general code model. When the formal idealized model replaced the formal continuum model, interactions between spheres and bounding surfaces were represented by simple mechanisms which came into existence at any time that a sphere became close enough to another sphere or to a boundary. In the general formulation of a DEM model, these simple representations are the same for every interaction. If there is a Hookean spring in the mechanism, then its spring constant is assumed to be invariant over time, location, and the history of each sphere or boundary component. This enforces a separation of scales in the model, which makes it impossible for the model to represent small-scale differences in grain geometry which can affect interactions.  

Uncertainties arising from smaller-scale inhomogeneity can be represented by choosing coefficients randomly for each interaction. If a Hookean spring is part of an interaction mechanism, its spring coefficient could be sampled from some chosen probability distribution whenever an interaction occurs. This coefficient would then be used for the duration of the interaction. The random sampling could be done independently, or else made dependent on time or on the interaction histories of the colliding spheres. By having the variability of the coefficient distribution increase over time, the  effects of damage could be simulated. 

If probabilistic models for unpredictability and interaction uncertainty are added to the general code model, something very different is obtained from the original purely deterministic model. Conditional on any particular set of sampled coefficient and initial condition values, the general code model is still producing what might be approximate solutions to a system of ordinary differential equations. If the approximation is good, the formal solution to the system of the ODEs is defining a trajectory in a high-dimensional state space from the initial state to the final disordered packed state, assuming that the system has a unique solution. The probabilistic models for the unpredictable initial conditions then create random samples of approximated trajectories which lead to a random sample of simulated disordered packed states. Use of uncertainty modelling changes the constraints that define individual trajectories, with the intention of enabling more realistic trajectories to be approximated. Since the probabilistically augmented general code model approximates random samples from an ensemble of trajectories, it will be considered to be an example of a general \textit{random trajectory (RT)} code model. 

\subsection*{Random Trajectory Modelling}

Random trajectory modelling is an approach to modelling complex, unpredictable phenomena which can be used whenever code-based simulation models for individual realizations of the phenomena are available. It can be applied far beyond models for granular flows. It can be applied to any phenomenon whose important aspects cannot be described by a simple formal mathematical model and the only feasible models are simple code models which simulate individual realizations by representing their emergence from many strongly-interacting smaller-scale elements. 

When discussing RT modelling, the phenomenon of interest will be some phenomenon in the physical world whose individual realizations are not simple, but whose realizations can be created by some relatively simple procedure that permits repeated observations of the phenomenon. It will be assumed that these phenomena are the result of laboratory procedures undertaken in controlled conditions, rather than passive observations of a natural phenomenon outside of the laboratory. Phenomena not produced in the laboratory may not be repeatedly observable, which would make validation of models for them much more challenging.  

\subsubsection*{Elements, Mesoscales, and Microscales}

Random trajectory models are multiscale models. They require that each realization of the phenomenon of interest be describable as some collection of well-defined elements whose individual states can each be described by a small number of variables. It is also necessary that elements have non-trivial interactions which can be represented by relatively simple models. If a useful simple interaction model is chosen, then a specified RT code model may be able to produce a useful representation of a realization of the phenomenon of interest. 

Elements can be defined by identifiable physical entities, such as the grains of a powder or the different material components of a fibre-reinforced composite material. They can also be defined formally as partitions of a continuum, but examples of this type of element will be deferred until after further discussion of RT modelling using  physical elements. 

Physical elements are almost always defined on a mesoscale, and not on the smallest observable or conceivable scale. Physically defined elements are considered to be composed of smaller scale entities, but these entities are not represented in the modelling.  This may require not representing observable physical entities below a fixed size, or ignoring interactions of elements with other non-modelled smaller-scale physical entities. This kind of idealization is necessary to build a usable general RT code model. To justify this abstraction, it is assumed that entities whose size lies below some limit have no significant impact on the emergence of features of interest for the phenomenon. In the modelling of the powder flow, ignoring debris and small-scale damage are examples of this form of separation-of-scale abstraction.

When modelling concrete, separation of scale assumptions are risky. It is not possible to define a smallest scale for the material that is practical for representative modelling of all larger structure. Concrete consists of aggregate (rocks) of many different sizes, throughly mixed and held together by a matrix of cement. The cement itself is a composite of crystals and gels which also has a disordered structure, so that if cross-sections of 1$cm^2$, 1$mm^2$, or 1$\mu m^2$ are examined with suitable technology, the images all show a similar state of complex disorder \cite{myc}. To make a usable simulated model of a concrete specimen, it is necessary to decide that only rocks of a certain minimum size are to be modelled as elements, and all smaller aggregate is to be considered to be a uniform phase with the cement. A random trajectory model can be used to generate a packing to represent the spatial arrangement of aggregate, and then statistics which estimate physical properties of the uniform binding phase can be calculated. This kind of model will be best suited to concrete made from large rocks and cement, in which intermediate grades of aggregate are omitted. This is a material which may have very different properties than any concrete used in applications. 

\subsubsection*{Three Kinds of Non-Simplicity}

RT code models can simply and explicitly represent three distinct ways in which a phenomenon may fail to be simple. They can satisfy Anderson's principle that `more is different' \cite{anderson:1972} in a way that is distinct from models for complex phenomena used in statistical physics. 

A phenomenon has realizations which are \textit{disordered} when there is no known symmetry or form of order that allows a useful and mathematically simple description to be made of any one observed realization of the phenomenon. Instead, the only way to describe a single experimental realization is to break that realization down into smaller scale elements and to describe the state of each element. If a small number of attributes for each element are sufficient to construct a useful description of that realization, then RT modelling of the phenomenon may be possible. Building models of this type is necessary when the disordered states are of the kinds which cannot be usefully represented by formal methods from statistical physics \cite{ziman}.

\textit{Unpredictability} arises when realizations produced by experiments are sensitive to physical conditions at the start of the experiment, and these conditions cannot be controlled. No matter how much effort is made to make the initial state of the elements the same between experimental trials, small differences in initial states are taken to be responsible for how the states of the elements evolve differently over time in every different experimental realization of the phenomenon. These small differences also produce between-realization variability in interesting aspects of the phenomenon.

\textit{Uncertainty} results from no two interactions between elements ever occurring in exactly the same way. These differences are generally unobservable \textit{in situ} as an experimental realization of the phenomenon evolves over time. 

\subsubsection*{Modelling Disorder}

To model a single realization of a phenomenon that is difficult to describe, it is necessary to define a collection of elements from which the realization will emerge. If the elements are physical, this requires specifying which smaller-scale entities in the physical world are to be represented. This may also involve making careful choices of the experimental phenomenon to model, in order to minimize the impact of the abstraction and idealization needed to build a usable model for the chosen phenomenon. 

The modelling of elements requires choosing a minimum size of the physical elements in the phenomenon which will be represented. It also requires specifying how many attributes of the physical elements will be represented in the  model. This requires using as few attributes as possible for each element, and always results in many aspects of the physical element being ignored in order to be able to create a usable model. The state of a realization of a phenomenon with $N$ elements and $k_i$ attributes for the $i^{th}$ element at a fixed time will be represented by a point in a $N^*=\sum_{i=1}^N k_i$-dimensional state space. For the packing example, a flow of $N=1000$ monosized spheres requires 9 attributes per sphere and state space of dimension $N^*=9000$. For many phenomena, state spaces with several orders of magnitude higher dimension will be required.

The representation of disorder requires a being able to define trajectories in the state space for two reasons. It may be that the time evolution of disordered states is the phenomenon of interest, and so the trajectory needs to be modelled. When only a static disordered state is needed, it will be necessary to define a trajectory that eventually leads to the right kind of static disordered state. In the case of a static sphere packing, a single realization of the packing can be represented by a single point in high-dimensional state space. Points in the state space that can represent sphere packings are extremely rare, and they cannot be identified by any simple rule or found by randomly selecting points in the state space until a feasible point appears. Some evolutionary process starting from a simple initial state is needed that can reach a point that might represent a packing in the state space, even if the details of the trajectory to that feasible point are of no interest. 

There are three general types of trajectory through the phase space which are relevant to disordered states. When the process of reaching a disordered state is important, there is a \textit{conceptual path} for an experimental realization of the process that constructs that disordered state. The potentially observable attributes of the physical elements in a single experimental realization of a disordered state can be imagined to define a continuous curve over time from the initial state of the elements to their final disordered state. It is almost never possible to observe all of the attributes of every element as it evolves in time. When it is possible to make such observations, those observations are taken at fixed times (or over time intervals) and the attributes are observed with error. As a consequence, the conceptual path of any one experimental realization can be imagined and never exactly observed. Conceptual paths are the kind of trajectories that any model for disorder is intended to simulate. 

The other two types of trajectory arise from models. When the trajectory is formally modelled by a system of differential equations that has a unique solution for any one set of initial conditions, then this formal solution can almost never be expressed analytically and so is also imagined, not observable, and not directly useful. The only trajectory that a researcher can work with directly arises from an implemented code model that might produce approximations to those formal solutions at a finite set of time points.

In the case of a model for the formation of a packing of 1000 monosized spheres, each experimental realization would be associated with an unobservable conceptual path through a 9000-dimensional state space. A fitted formal model for the experimental process would also define a trajectory through that space, but it would be unobservable due to the lack of an analytic expression of that formal solution. Given a set of initial conditions, a fitted code model might be able to produce a finite sequence of points approximating that formal solution at a sequence of times. 

Representation of disordered states is complicated by two aspects of using a dynamical system to represent many interacting entities. The first is that there may not be a unique representation of any one experimental realization of the phenomenon in the state space. If there is no general rule for distinguishing and uniquely identifying the elements a disordered initial state, in the worst case there could be up to $N!$ different ways to label the $N$ elements in a model, and $N!$ different trajectories in $\mathbb{R}^N$ capable of representing the same realization. Even if this were not an issue, the method of modelling the disordered state requires using some form of algorithm for trajectory-building which is sensitive to initial conditions. Since the exact initial state of an experimental realization is generally not observable, there is no way to model a single realization with a single model trajectory. If an attempt to model a single experimental realization differs from what was observed, there will be no way to determine if the differences arise from mis-specification of the initial conditions, mis-specification of the model parameters, or the the fundamental inability of the idealized and abstracted model to be able to represent the realization. 

A discrete formal model for a trajectory through disordered states is of no use unless it is augmented so that the model can sample from an ensemble of trajectories. This is necessary for the model to be able represent the non-replicability of experimental realizations. This can be achieved if small changes to the initial conditions of the disorder model produce many different disordered structures of the same general kind. The set of all trajectories with viable initial conditions then becomes the model ensemble. In the absence of interaction uncertainty, it is important that the model trajectories be uniquely defined for each set of initial conditions and by the set of constraints on how that trajectory can evolve over time. It is also necessary that these model trajectories can be reproduced on different computers given only the initial conditions and the constraints on how the path can evolve. Each particular trajectory model needs to have adjustable parameters so that they can be fit to observations, ideally via parameters whose meaning might be connected to some useful theoretical narrative. This is needed when a theoretically-based formal model could apply to many distinct yet similar experimental phenomena. It is also needed because it is very unlikely that the first choice of parameter values for a new model will result in useful output.

There is an alternative to using a code-based model on a high dimensional state space to represent the disordered realizations of an experimental  phenomenon. This would involve using a physical model: another replicable experimental phenomenon that is considered to be similar enough to the phenomenon of interest to be useful, but also is more easily observable and manipulable than the phenomenon of interest. Bernal and Finney's work on monosized sphere packings of steel balls \cite{bernal:1959,finn:1970,finn:1970b} involved studying a disordered state that might have the same kind of disorder as atoms in a liquid or a disordered solid, so as to gain insights into those not directly observable phenomena.  

\subsubsection*{Modelling Unpredictability}

To represent the inability to replicate experimental realizations, it is necessary to be able to randomly sample the initial conditions for the dynamical system that represents the disorder in any one realization. This is necessary whenever important emergent aspects of the experimental phenomenon remain unpredictable. 

Sampling initial conditions for a random trajectory model requires being able to find a useful ensemble of initial conditions. 
This can be difficult if the initial state of the system is itself disordered. In the worst case, the initial states may be of the same kind that the model is intended to create. This kind of circularity can be avoided if initial states are used which are far simpler than any associated with experimental phenomenon, and a strong case can be made for the excessive simplicity of those conditions having no impact on the utility of the model. 

The most simple way to simulate initial disordered states is to use a physical model. If an initial state is a static packing of monosized spheres, then it may be possible to use remote sensing to identify the locations of sphere centers within an experimental realization of an initial packing. If no technology exists to find these centre locations with minimal error in an experimental realization, then it may be necessary to produce realizations of packings of some other kind of spheres from which positional information can be found. If the spheres of the imageable material are formed into packings by the same experimental procedure as the spheres in the packing of interest, images of $n$ replicates may be used as a random sample of size $n$ of initial states. If the spheres of a different material do not form packings in the same way as those in the packing of interest, then this approach may result in fitted RT code models which cannot produce a sample of approximate trajectories similar to those of the experimental phenomenon of interest. 

If no physical model is available, then it may be necessary to use a different code model to generate the initial states. This can be done using a model with the same theoretical basis as the RT model being constructed, or by using a model that has no connection to the theory but which can produce similar disordered states.

If a model with the same theoretical basis is to be used, then it is necessary to find some part of the trajectory of the dynamical system model for a disordered state which is easier to simulate than are the initial states. In the case of pouring one packing of spheres from one container into another, a model for the whole process should begin from a packing of spheres. If the pouring process occurs over a long distance, the spheres may disperse while falling into a state where no spheres are touching each other. If the modelling is started at this time point, it may be possible to simulate the initial positions by simulating from some probabilistic spatial point process whose realizations have points sufficiently far from each another that no two spheres are in contact. The velocities of the spheres could be assumed to all be unidirectionally downward, although this could be an oversimplification. The velocity distributions of the falling physical spheres may be related to their histories since being poured from the packed state. If these distributions were mis-specified, again this approach may result in models which cannot produce a sample of model trajectories similar to a sample of experimental ones. 

In some cases, it may be possible to generate disordered states by methods which have no basis in any physical theory that could explain how those states came to be. If a packing of monosized spheres is needed for an initial state, codes have been developed which can re-arrange any initial disordered pattern of $N$ points into an arrangement approximating the centres of a packing of $N$ monosized spheres \cite{vold:1964,vb:1972,jt:1985,barg:1991,lochmann}. These algorithms can be initialized by random selections of $N$ points from a uniform distribution over a region, enabling random samples of packings of any size to be easily produced. Since these algorithms have no formal connection to the mechanics of physical sphere packing formation, it would be necessary to hope that these algorithms produce packings which are indistinguishable from those that could be produced by an experimental process in the physical world.

\subsubsection*{Uncertainty in Interaction Models}

The complexity of a realization an of experimental phenomenon results from each element interacting with many others. In the simplest theoretically-based formal models for a single disordered state, each interaction between elements will be represented in the same way. In an experimental realization, each interaction between elements may occur differently due to many small-scale differences between elements which are not accounted for in an idealized and abstracted model. 

To model each interaction differently, the general code model for a single disordered state needs to be modified. If there are simple models for interactions which contain constants whose values need to be specified, then those values could be randomly selected from some chosen probability distribution when the interaction begins. This random choice could be made from the same distribution each time, which need not be Gaussian. It could also be made from a distribution which changes based on when the interaction occurs, based on where the interaction occurs, based on other elemental aspects at the time when the interaction occurs, or based on the history of the elements involved. Such distributions could be applied to some or all coefficients present in the mechanistic model of the interaction. 

Adding models for interaction uncertainty adds three more challenges to the overall RT modelling process. 

Adding probability distributions for constants in an interaction introduces more coefficients that need to be fit. Instead of a single spring constant to be fit, it may be necessary to fit two or more parameters for the distribution of the constant. Since RT models work best when there are as few coefficients to be fit as possible, this limits the number of simple probabilistic interaction models that can be used.

When choosing the family of distributions needed for the coefficients, there are generally no theoretical or empirical arguments for choosing the family to use. There is no obvious reason to use a Gaussian distribution, and a skewed or long-tailed distribution may be more useful. Several different families of distributions may need to be tried before a useful RT model can be constructed. 

The third difficulty arises from the uncertainty model being added to the general code model for disorder, which is based on a deterministic dynamical system. As long as the code still runs and produces output that appears to represent the right kind of disorder, adding an uncertainty model to the code model can potentially improve the utility of the random trajectory model as a whole by making it more realistic. Without the uncertainty model, the general code model is the approximate solution for a formal dynamical system, specified by a system of differential equations. When uncertainty models are added to the code, it may not be possible to define a formal model whose solution the code is approximating. 

\subsubsection*{Basic RT Modelling of Complexity}

RT modelling of a complex experimental phenomenon begins with choosing a way to think of realizations of the phenomenon as being emergent from the relatively simple interactions of many elements. It also involves choosing a relatively small number of elemental attributes that are to be observed and modelled, and requires choosing a state space in which those attributes can be represented. 

A formal model for the disorder in a single realization is then chosen, generally as a parameterized family of dynamical systems that constrain which points in the state space are viable representations, and by defining trajectories which can connect such points as the system evolves over time. The dynamical system may be a system of differential equations with solutions continuous in time, but could also be a discrete dynamical system like a cellular automaton. If the dynamical system is not already discrete and represented in code, it will need to inspire a code model which might be able to approximate its trajectories. The code model will need to be sensitive to initial conditions. If interactions are thought to affected by attributes that are not represented in the model, a model for interaction uncertainty may be added. 

The model for the individual trajectories must be accompanied by some process which can sample initial states for the code model. These sampling processes will be probabilistic, and so the fitted RT models will be randomly sampling time series which might be able to approximate conceptual paths arising from the ensemble of possible realizations of the experimental phenomenon. 

\subsubsection*{RT Modelling with Non-Physical Elements}

The initial example of RT modelling was based on an example in which the elements represented well-defined physical entities in realizations of an experimental phenomenon of interest. If an experimental phenomenon occurs in a medium which is modelled as a continuum using partial differential equations (PDEs), it is possible to use RT modelling as long as the disorder of an individual experimental realization of the phenomenon is modelled by a system of PDEs which is sensitive to initial conditions. This requires allowing elements in the RT model to be arbitrarily defined by the modeller or by software, but extends the RT modelling framework to many simulation models based on numerically solving systems of PDEs. This extension provides a way to clearly discuss essential problems that arise when fitting and validating PDE-based models for complex experimental phenomena. 

Disordered realizations of complex experimental phenomena can be modelled by assuming that the phenomenon occurs in some form of continuum, rather than by means of well-defined physical elements interacting. Examples of these phenomena include turbulent fluid flow \cite{durbin:2017}, complex patterns of stress and strain within solid bodies \cite{majmudar:2005,dix:2022}, and many other pattern-forming phenomena \cite{ball:1999} including Rayleigh-B{\'{e}}nard convection \cite{debruyn:1996}, Couette-Taylor phenomena \cite{fardin:2014}, patterns formed in chemical reactions \cite{epstein:1996}, patterns emerging from vibrated granular matter \cite{melo:1995}, and biological patterns seen in animal skins and in patterns formed by vegetation in arid environments \cite{murray:2003}. While all of these phenomena are known to result from the interactions of well-defined physical elements, the use of continuum modelling allows for the development of relatively simple and useful models for these phenomena that would not be available if physical elements were used. 

Continuum models for disordered realizations are systems of PDEs which describe how some small collection of attributes of a continuous medium are constrained to change over time and space within a realization. These equations are generally based on theoretical narratives which require the conservation of energy, momentum, and mass, but the constraints are based on constitutive equations which are often empirical or heuristic. The construction of these equations involves abstraction and idealization in order to produce a model which is simple enough to be useful. These equations will have adjustable coefficients which may need to be fit, and tend to be non-linear and sensitive to initial conditions. 

Formal studies of patterns generated from a model based on a system of PDEs can sometimes be undertaken using analytical methods from the qualitative theory of PDEs \cite{cross,hoyle,getling,ci}. These analyses can produce useful insights about what properties formal solutions to the PDEs may have. Fitting these models to data or validating their applicability to the phenomenon requires that approximate solutions to the PDEs be found which can be compared to observed realizations of the phenomenon in some way. 

The extension of the RT modelling framework to complex phenomena in continuous media requires specifying the models for disorder and for unpredictability. The systems of PDEs constitute a general formal model for the disorder of a single realization when solved for a fixed set of coefficients and a fixed set of initial conditions. The chosen PDE solver code for the system of PDEs plays the role of the general code model which can provide an approximate solution. When the general code model is combined with probabilistic selection of initial conditions, a general RT code model can be constructed which can potentially represent the unpredictability of those disordered patterns as they are observed in repeated realizations of the phenomenon. It is also a random trajectory model, but the definition of those trajectories is more complicated than in RT models based on physical elements.

When a continuum model is used, the formal solution to the system of PDEs defines the value of $k$ attributes at every point in a finite region where the phenomenon is being represented. The solution is defined at an uncountable number of spatial locations, so there is no finite dimensional state space associated with it. When a solver is constructed or applied by a researcher to the system of differential equations, the solver initially partitions the region into $N$ volume elements. If this partition remains fixed over time as the solver builds the approximate solution, then at each time point the approximation to the solution can be represented as a point in an $Nk-$dimensional state space. The system of PDEs is replaced with a system of partial difference equations, but the need to satisfy spatial constraints among the elements will mean that the details of how the approximations are calculated are much more complex than when only time derivatives are involved. 

There is no unique way to define the $N$ volume elements used in approximating the formal solution to a system of PDEs. Definition of the elements is based on striking a balance between approximation accuracy and limitations on time and computing resources. For any one formal PDE model, every different kind of solver will produce a distinct general code model which will produce different outputs for any one set of initial conditions and coefficient values. 

Since the elements are user-defined, problems arising from the arbitrary labelling of physical elements do not occur as long as the same elements are used throughout the construction of the approximate solution. Most solvers redefine the elements between time steps in a way that is history-dependent, so that the total number and meaning of the elements change over time in a way which is different for every choice of initial condition and coefficients. While these solvers can be viewed as still producing a time series of points in a high-dimensional state space based on a given set of initial conditions, for different initial conditions those paths will be incomparable with each other. When the same elements are used at all time points, it is possible to imagine each solver solution as approximating a continuous path in an $Nk-$dimensional state space which represents how the spatial approximation evolves in continuous time. 

Another issue arising from the arbitrary nature of elements in solver models is that there is no clear way to compare raw solver output to observations of realizations of the experimental phenomenon. When elements are physically defined, it is possible to imagine a conceptual path for each observed realization by imagining how the $k$ attributes of the physical elements in a realization would change over time from their initial state. When the phenomenon is assumed to occur in a continuous medium, there needs to be some way to discretize the observations of realizations as they occur in a continuous medium. This may require use of a reliable model for the behaviour of the fluid which may be very similar to the RT model that is being constructed. It would also only be possible to make such comparisons when the same elements were used in every time step and for all sets of initial conditions.

When considering PDE-based continuum models in the random trajectory context, it will be assumed that the solvers define elements which remain unchanged over all time steps and initial conditions. When this is done, it will be possible to clearly show important difficulties which arise when attempting to fit or validate these models. Since output from general solver models is more complex than what is found when the elements remain the same, it is expected that all difficulties found in the fixed element case will also be found in more general solver-based RT models. The proposed methods for fitting code models and validating the fitted code models developed for RT models with physical elements can be applied to both kinds of solver-based models. 

\subsection*{What utility might a fitted RT code model have?}

When coefficient values are fit to a general RT code model, it becomes the first mathematical or computational entity that can produce anything that can be compared to experimental observations. In order to understand model fitting and validation in these contexts, it is necessary to understand the limitations of the general RT code model.


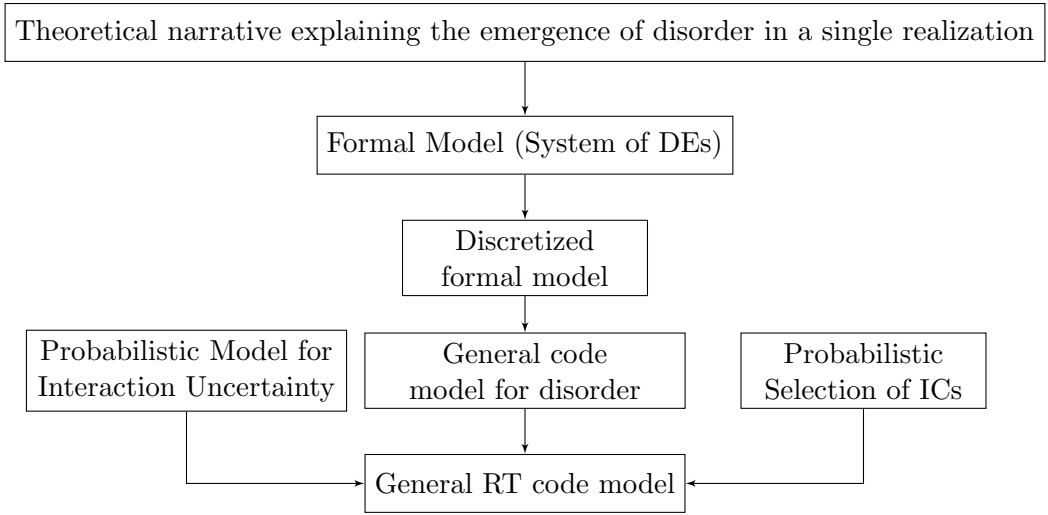
\begin{figure}
\centering
\begin{tikzpicture}

\node [process] at (0,0) (narr) {Theoretical narrative explaining the emergence of disorder in a single realization};
\node [process] at (0,-1.5) (de) {Formal Model (System of DEs)};
\node [process, text width=3cm] at (0,-3) (dfm) {Discretized formal model};
\node [process, text width=4cm] at (0,-4.5) (gcmd) {General code model for disorder};
\node [process, text width=3cm] at (4.5,-4.5) (pr1) {Probabilistic Selection of ICs};
\node [process, text width=4cm] at (-4.5,-4.5) (pr2) {Probabilistic Model for Interaction Uncertainty};
\node [process, ] at (0,-6) (gcm) {General RT code model};

\path [connector] (narr) -- (de);
\path [connector] (de) -- (dfm);
\path [connector] (dfm) -- (gcmd);
\path [connector] (gcmd) -- (gcm);
\path [connector] (pr1) |- (gcm);
\path [connector] (pr2) |- (gcm);

\end{tikzpicture}

\caption{Stages in the construction of a general RT code model}
\label{construct}
\end{figure}

Any general RT code model has a history that must be examined critically when assessing its potential utility. Figure 1 shows the most basic stages in the construction of a general RT model. Each stage is constructed by some combination of idealization, abstraction, and approximation that is necessary in order to eventually be able to produce code that can simulate some aspects of a sample of realizations. At each transition between stages, there is the possibility that the final model will become causally disconnected from the original narrative that inspired it. Even if the narrative explanation is a comprehensive and useful explanation for the emergence of the complex experimental phenomenon itself, excessive idealization and abstraction may produce a formal model which cannot represent that narrative. If the formal model is representative and well-posed, then the discretization, approximation, and details of coding may result in a code model which is not an approximation to the solution to formal model. 

In the best case, the general RT code model will have the capacity to faithfully represent how important aspects of a realization of a phenomenon emerge from the interactions of its constituent elements. Nothing would be gained by adding representation of smaller scale interactions or adding additional attributed to each element. Meanings of coefficients in the general RT code model would have the same meaning as they were intended to have in the formal model. The approximations used would differ from what they approximate by trivial errors which could be ignored. The uncertainty component in the model would select initial conditions in a way that was statistically indistinguishable from how the initial conditions of many replicates of the experimental phenomenon are distributed. There would be a single set of meaningful coefficient values which would yield a specified code model which could produce samples of approximations to formal model solutions which would come from an ensemble of trajectories which are indistinguishable by any means of statistical inference from the ensemble of conceptual paths arising from experimental realizations. The fitting procedure would estimate those coefficient values. This kind of general RT code model would then be a \textit{elementally representative} model for realizations of the phenomenon. 

It is not clear that any output-producing general RT code model can be elementally representative. If the model does not include attributes related to the emergence of the phenomenon in the physical world or mis-represents them in the formal model, if the seemingly-useful code model is constructed to approximate a formal solution that does not exist, if the solver approximations fail to approximate the formal model solution, or if the distributions for probabilistic components of an RT model are highly unrepresentative, then the general RT code model will not be able to represent the mechanism that causes the experimental phenomenon. If this were the case, improperly validated fitted model output could serve as overvalued evidence in favour of explanations which are fundamentally wrong. If there is any risk of this kind of failure happening, then the most important aspect of validation will be providing strong evidence that the general RT code model is elementally representative. If no such evidence is available, it is necessary to assume that one or more of these failures have happened, and so the general RT code model is \textit{elementally false}. An elementally false RT model cannot directly provide evidence for any explanation that it was constructed to represent. 

If a general RT code model is elementally false, then it may still be very useful for simulating some aspects of a complex phenomenon. A \textit{simplifying statistic} is some numerical value which is calculated from data observed on an experimental or simulated realization of a phenomenon. In an experimental setting, these data may be collected via instruments directly, or from images of an experimental realization. Simplifying statistics are random in that their values for a realization at any particular time cannot be predicted before observing a realization of the phenomenon. In spite of the fitted RT code model not being able to represent the physical mechanism responsible for the emergence of important aspects of the experimental phenomenon, there may be some way to fit an elementally false RT code model so that the joint distribution of simulated simplifying statistics is indistinguishable from the joint distribution of observed simplifying statistics by means of statistical inference. The use of fitted RT code models may make it possible to emulate joint distributions which cannot be represented well by standard multivariate probability distributions or by copula-based methods. This kind of utility may be available since a fitted elementally false general RT code model may be least unrepresentative of the phenomenon than all other models that are available. If the fitted RT code model can emulate the joint distribution of some collection of simplifying statistics, then it will be possible to make estimates of means, variances, correlations, quantiles, and conditional probabilities based on simulated samples of the simplifying statistics. Emulation is a necessary condition of any model used to justify an explanation, and so being able to find evidence of the capacity to emulate will be a necessary part of validation of fitted RT code models used for prediction or as evidence for explanations. 

The possibility that a general RT model can be misrepresent the cause of the phenomenon but still useful is of core importance, since the general RT code model cannot be fit or validated based on its raw output. Even if there were no difficulties with elemental identification, all samples of time series of approximations whose values lie in state spaces with thousands or millions of dimensions will be too small to be plausibly representative of the ensemble from which they were sampled. In order to compare model output with observations in order to fit coefficients, the fitting must be done on the basis of a relatively small number of simplifying statistics which focus attention on aspects of the phenomenon which researchers find useful or interesting. This type of fitting produces a \textit{fitted restricted RT code model}, whose output consists only of the simulated observations of the simplifying statistics used to fit it. Once coefficient estimates are obtained from a fitted RT code model, those estimates can be applied to a general RT code model to create a fitted general RT code model. This model can then generate approximate trajectories from which images can be created or from which distributions of simplifying statistics not used in the fitting process can be simulated. 

\pagebreak

The process of mapping from a state space with thousands or millions of dimensions down to generally less than 10 simplifying statistics discards much of the information present in the general RT code model output \footnote{The choice of `simplifying' rather than 'summarizing' for the designation of these statistics focuses on the loss of information that occurs when they are used, and on the way that their use imposes simplicity on the model fitting process that makes fitting possible.}. If those simplifying statistics inherit their probability distributions in ways that are insensitive to details of elemental interactions, then elemental falsehood may not affect the ability of those fitted restricted RT code models to simulate observations of the chosen simplifying statistics. The coefficient values providing the best fit may vary greatly when the simplifying statistics are changed, so a restricted RT model fit to one set of simplifying statistics should not be expected to be able to accurately emulate the distribution of any other simplifying statistics that were not used in the fitting process. 

In the absence of a strong argument that the general RT code model is elementally representative, a fitted restricted RT code model may only have the capacity to emulate the joint distributions of the simplifying statistics used to fit it. Emulation is also a property of any well-fit elementally representative RT code model, but the capacity to emulate is not sufficient to provide evidence that a fitted RT code model is elementally representative. 

Fitting and validation procedures for RT code models need to be based on the assumption that the general RT code model is elementally false, but still perhaps capable of emulating the joint distribution of a collection of simplifying statistics. The fitted restricted RT code model needs to demonstrate its validity by means of new observations, rather than by its consistency with accepted theoretical explanatory narratives. 

\section*{Fitting a restricted RT code model}

Fitting a restricted RT code model requires choosing coefficients which enable the specified model to produce useful output. The process of fitting requires the use of random samples both of specified general RT model output and of realizations of the phenomenon. The fitting process is a form of statistical inference, and so the known limitations of all stochastic fitting processes will complicate any validation of the fitted RT code models. 

\subsection*{Is it necessary to use random samples?}

If a complex phenomenon has simplifying statistics whose values are not invariant between experimental realizations, then the variability and dependence of those statistics can only be represented by a model that can simulate random samples of those statistics. The general RT code model must contain stochastic unpredictability or uncertainty components in order to generate random samples of simulated realizations from which samples of simplifying statistics can be collected. All samples must be random in order to have a reasonable chance of being representative of the ensembles from which they are sampled. There is no guarantee that any one random sample will be representative of its ensemble, but random samples will be free from the bias that can emerge when samples are constructed by expert personal judgment or through the careless assembly of easily available data.

The quality of the random sample depends on its size $n$. For each set of candidate values for the coefficients, a specified RT code model needs to produce $n$ independent simulations from which the simplifying statistics can be constructed. Increasing the number of elements $N$ within a realization does not improve the quality of a sample of fixed size $n$, but instead may fundamentally change the ensemble being sampled by the model and also change the phenomenon itself. 

\subsection*{Randomly sampling realizations}

While a probabilistic model can provide random samples as large as computing resources allow, random sampling of experimental realizations needs to be undertaken implicitly through the careful construction of an experimental procedure that establishes how realizations of the phenomenon are to be produced. Ideally, the phenomenon must have enough simple aspects that its realizations can be distinguished from realizations of other phenomena, and that other researchers can produce their own realizations of the phenomenon. If the phenomenon of interest is very sensitive to small changes in how experimental realizations are produced, then this may result in modelling results not being replicable because every fitted model is being fit on data arising from fundamentally different phenomena. 

In the case of sphere packing production, the best data from model fitting would come from a procedure which specifies the kind of spheres to be used (material, shape characteristics, size, and source), the kind of containers to be used (specified to the same degree), the ways in which the containers and spheres are to be prepared, the way in which the initial container is to be filled, and the way in which the spheres are to be poured from one container to the other. The last step might be instructions for a person, or for the construction and operation of a mechanical pouring apparatus. There may be instructions on climate control, or on how and when the data on final sphere location is to be collected once the spheres have been poured. If researchers follow the procedure exactly, it should result in more observations of a phenomenon which should arise from the same physical mechanism as the original observations, and which should possess the same kind of disorder within individual realizations and the same kind of unpredictability between realizations. If the procedure is communicated well, it would allow other researchers to create essentially the same pouring process in different locations at different times with different apparatus. If such a procedure exists and it is reasonable to assume that uncontrolled aspects of the phenomenon vary randomly from realization to realization in the same way each time a realization is created, then a collection of realizations created by following the procedure can be considered to be a random sample from all experimental realizations that could ever be produced by following that procedure. 

Requiring that a phenomenon be defined by a specific, detailed experimental procedure greatly limits the kind of phenomena which can be modelled. This requirement is important in initial stages of modelling since complex phenomena have no obligation to be easily replicable, and should not be expected to be easily replicable. Even in the case of having a well-defined procedure, if the procedure isn't followed or excludes control of important aspects of the phenomenon, then it may not be possible to obtain even a single well-defined sample. If the spheres in a pouring experiment are not cleaned between uses, build-ups of dust and swarf on the spheres may change the phenomenon over time. If the same spheres and containers are used many times, the pouring process may sufficiently machine the spheres and the container so that later experiments involve spheres which are much less spherical and containers which are much less cylindrical than they were in the initial realizations. If the procedure is poorly written or perceived as being too strict, later researchers could end up creating new phenomena based on how they interpret the instructions or fail to follow them. 

If a phenomenon is created in a way that changes uncontrollably over time, it will sample from multiple ensembles of phenomena in a way that may not be replicable. If a different lab claims to follow a procedure but creates samples from a different ensemble of phenomena, then combining the samples or comparing models fit to the samples would yield misleading evidence. Model validation processes must include analyses to look for evidence of this kind of problem. It is possible to look for evidence of time dependence in the the values of each simplifying statistic observed on many different samples, and also to look for evidence of major differences in the distributions of the simplifying statistics produced by different labs, by different personnel within the same lab, or on different apparatus. Some of the analyses used to validate fitted RT code models may also be useful for making comparisons between observations of simplifying statistics which arise from different samples of realizations. 

Many phenomena of interest are observed outside of the laboratory and their occurrence is identified by expert judgment. There is a strong risk that realizations of a phenomenon defined by expert judgment might emerge from multiple distinct physical mechanisms which are not easily observable. It is also the case that a complete list of every observation of a  non-experimental phenomenon in the physical world may be far from being representative even when the phenomenon is well-characterized. Whether or not an instance of the phenomenon is observed may depend on whether or not it occurs near a human observer, and on there being some reliable way to record that it occurred. Lists of available observations of instances of a well-characterized non-experimental phenomenon should not be considered to be random samples unless a very strong case can be made that they are likely to be representative. 

\subsection*{Choosing the coefficients to fit}

A restricted RT code model will have many coefficients to fit. These will include the parameters of probability distributions used in the uncertainty and unpredictability models, but most coefficients will come from the formal RT code model for disorder. 

If a restricted RT code model is assumed to be elementally false, then the original meanings of the RT code model coefficients in the formal will be lost. Without any clear meaning attributable to the coefficients, all of them should be fit from data. It is usually impossible to achieve this. Since fitting models of this kind requires generating $n$ simulations of a specified code model for every candidate set of coefficients, there are constraints on how many candidate sets can be tried. If too many coefficients need to be fit, there may not be enough resources to undertake a search through enough candidate sets to find a model that fits well. Although all coefficients in the three components of a restricted RT code model should be fit, it is necessary to make choices about which ones must be fit using the data and which ones can be fit by other means.

The number of coefficients to be fit can be reduced by idealization and abstraction when setting up the formal RT code model for disorder. In the case of packing simulation, some of the many forces acting on the spheres could be omitted. If there are several models available for a particular force, the one with the least number of coefficients could be chosen. Adopting this strategy increases the likelihood that the general RT code model will be elementally false.

One way to choose coefficients to fit is to examine all of the components of the formal model and sort them according to robustness. For a model for a sphere packing created near the Earth's surface, the gravitational force (in newtons) acting at the centre of a sphere would be $9.8m$, where $m$ is the sphere's mass in grams. The force of gravity is law-like, so there would be no other reasonable way to express it in a formal model for disorder that had the capacity to be elementally representative. The formula for electrostatic force and its coefficients would be treated similarly. Expressions for sphere-fluid interactions are empirical and were developed based on experimental phenomena very different from a disordered arrangement of many spheres flowing through a turbulent fluid. These expressions might still be reliable in some cases, but there would be no justification for believing that they would apply unchanged when applied to modelling a novel phenomenon for the first time. The coefficients of the springs, dashpots, rollers, and other simple mechanisms that are used to build a heavily idealized model of the forces involved in collisions of spheres are the ones which are likely to be the least robust, and should have the highest priority for fitting. 

For phenomena represented by pattern-forming systems of PDEs, the core PDEs may or may not be law-like. While the Navier-Stokes equations could be considered law-like, other equations in pattern-forming models can be much more idealized. For those PDEs, coefficient fitting from data would be necessary. When the PDEs are law-like, there may be parameterized auxiliary conditions (such as velocity profiles in fluids) which are not law-like, but which are needed in order for there to be solutions to the formal model. Coefficients in these expressions would be best fit from data.

Coefficients which are parameters in probabilistic components of general RT code models also need to be fit from data. In the case of uncertainty components, the distributions are chosen for convenience and they affect the terms which tend to be of the most idealized and least robust kind in the disorder model. In the case of unpredictability, the distributions may not represent how experimenters implicitly choose initial conditions when executing experimental procedures.

When coefficients are not fit using observations of simplifying statistics, their values need to be chosen by some other defensible process.  One approach is to assume that the components of the formal model will be elementally representative, and to use generally accepted coefficient values which have been useful in models for similar experimental phenomena. This may involve using standard expressions for physical laws of force, or using experimental estimates of coefficients found from observations of different phenomena. Trying to estimate spring coefficient values for simple mechanized models of contacts between flowing spheres by means of experiments involving pairs of spheres brought together in a testing apparatus would be an example of this kind of fitting \cite{cavarretti:2011}. Since the resulting fitted general RT code model is likely to be elementally false, this approach risks fitting values which differ from those which would yield the most useful fitted model. Coefficient estimates could also be chosen by using values which have provided useful results in models of similar phenomena, even if those estimates contradict previously established useful experimental or theoretical results. Intuitive guesses are also an option, though one of last resort. 

When general RT code models are assumed to be elementally representative, it would be expected that data-based estimates of coefficients which are physical constants would be close to the accepted values of those physical constants. If the gravitational acceleration in a gravitational force term were to be fit from data via a restricted RT code model and it were found that a value very different from $9.8 ms^{-2}$ gave much better model performance, then this could serve as evidence for the model being elementally false \cite{yen}. 

\subsection*{Choosing the simplifying statistics}

Simplifying statistics are required both for estimating coefficients and for validating fitted RT code models. It is necessary to find a small collection of statistics which categorize and quantify aspects of the phenomenon which are important to the model user. The typical behaviour and variability of these statistics need to be sensitive to changes in the coefficients that need to be fit. No pair of these statistics should be highly correlated. The statistics need to be observable from experimental realizations by means of measurement instrumentation or some form of expert judgment. There also needs to be a model for these instruments or processes of expert judgment so that observations of the simplifying statistics can be constructed from specified RT code model output. These requirements are challenging to satisfy, since there may not be sufficient numbers of existing statistics for the particular phenomenon of interest which are both useful observable summaries and whose observations can be simulated. 

Many simplifying statistics are observed via examination of still or moving images captured from experimental realizations by researchers or image analysis software \cite{WY:2025} . Finding simulated observations of these statistics requires simulating still or moving images from specified RT code model output via a model of the imaging process. The simplest imaging process models are likely to simulate images which have less noise, and which may also lack artifacts which arise from the imaging process. The imaging process model may excessively idealize and abstract the imaging process to the point that the imaging process simulation is representing something fundamentally different from the imaging process. These differences may result in simulated samples of simplifying statistics being incomparable to samples of observed statistics for model fitting and evaluation. 

Some problems arising from the use of images can be avoided if it is possible to use numerical or categorical simplifying statistics which are observed via measuring instruments interacting with realizations of the phenomenon. If this can be done, then it also necessary to have a model which represents the interaction of the measuring instrument with a realization of the phenomenon to produce the numerical output, but which interacts with output from a specified RT code model. Again, if this instrumentation model has different biases and variability from the experimental instrument, it is possible that samples of the simulated and observed simplifying statistics may be incomparable.

In many cases, simplifying statistics which are easily observed from specified general RT code model output may prove to be unobservable in experimental realizations. If those statistics have the potential to be useful, this may provide incentives for developing new measurement methods. 

For summarizing statistics to be useful in model fitting and evaluation, they need to be able to summarize aspects of complex disorder in a single realization of the phenomenon. This requires that the statistics be able to simply describe aspects of emergent structure within a realization which are characteristic of all possible realizations of that phenomenon. These emergent structures are likely to involve many elements, yet may or may not be small with respect to the entire realization. There are no grounds for expecting that a single set of simplifying statistics which is useful for modelling one phenomenon would be useful for other phenomena unless strong experimental evidence supported such extensions. There needs to be constant development of new simplifying statistics that can potentially summarize not-yet-seen forms of disorder arising from new phenomena, and for improving the fitting and validation of RT code models of existing phenomena. 

When searching for simplifying statistics, it is essential to look beyond statistics that have been useful for modelling phenomena that can be well-represented by models from statistical physics. In those phenomena, elements interact only locally and pairwise. Their disordered spatial arrangements at any fixed time can be seen as a realization of a stationary spatial point process model from probability theory \cite{minlos,dvj:2003}. That kind of disorder is best summarized by simplifying statistics constructed by making local measurements on a lattice of observation points (or at every element), and then averaging those local measurements over all observation points (or elements). The inherent smoothing in simplifying statistics from statistical physics may annihilate any useful information about within-realization disorder that could be obtained from a single realization, and also may make it very hard to understand what features of a realization are contributing to the particular value of the statistic. 

\subsection*{Estimating the coefficients by statistical inference}

When fitting coefficients, it is necessary to understand the limitations of the fitting process. If it were possible to unendingly repeat the experiment that produces realizations of the phenomenon, then the collection of all observations of the $p$ simplifying statistics would be described by a multivariate probability distribution. If the simplifying statistics were not all spatial averages, then their joint distribution might  not be Gaussian. If simulated realizations could be produced indefinitely, then there would also be a limiting joint distribution on the ensemble of all simulated observations of the simplifying statistics. These distributions would be different for each specified set of coefficient values. The goal of estimation would be to find the set of coefficient values which makes the joint distribution of the simulated observations as close as possible to the joint distribution of the experimental observations. If the general RT code model is elementally false, there may be no joint distribution of simulated observations that could be similar to the distribution of experimental observations.

In practice, the fitting process needs to be based on one sample of $n$ experimental realizations which can be considered to be random and on random samples from specified RT models for some finite collection of coefficient value sets. The fitting process requires defining a metric which can be used to determine which simulation produces a sample that is closest in some sense to the experimental sample. There is no single correct choice for this metric, but the choice of metric will affect the outcome of the fitting procedure. One commonly chosen metric is the Wasserstein (or earth-mover's) metric \cite{bernton:2019}.

Candidate values can be chosen by expert judgment, but more systematic methods have the advantage of being able to try candidate sets which experts may misjudge to be not worthy of consideration. Systematic methods may be deterministic or probabilistic, but all methods require choosing some compact subset of the space of candidate coefficient values and a starting point within that subset. The choice of region and starting point can be based on assuming that the formal model for disorder is elementally representative, on values which have been useful in similar models for similar phenomena, or on expert judgment. There is no guarantee that the subset chosen will contain a good solution to the fitting problem. 

A deterministic method for selecting candidate coefficient values would be to look at every possible combination of $m$ different values for each of the $p$ simplifying statistics. Since that type of lattice structure would require samples of size $n$ from each of $m^p$ different candidate sets, trying to look at every combination may be impossible unless $m$ and $p$ are both small. 

When fitting probability models to data, likelihood-based methods are considered to be the best approach. Although restricted RT code models are probability models, they are code models which makes it impossible to use the standard methods of maximum likelihood fitting. By using adaptations of Markov chain Monte Carlo estimation, it may be possible to probabilistically choose many candidate sets and to use them to build a sample from a Bayesian posterior distribution for the coefficient values for the \textit{general} RT code model by means of an approximate likelihood. Approximate Bayesian Computation (ABC) \cite{beaumont:2019,drovandi:2022} requires choosing a prior distribution for the coefficients, choosing a metric for the distance between the simulated and observed samples, and choosing a value for how close the two samples should be in order to be considered for selection into the posterior sample. When the simplifying statistics are sufficient statistics for the coefficients of the general RT code model, choosing a measure of typical value from a sufficiently large ABC posterior sample of coefficients can be shown to be close to being a maximum likelihood estimate for the coefficients of the general RT code model. 

\subsection*{Meanings of Coefficient Estimates}

Once coefficient estimates have been found by some algorithm, it is essential to recognize that the estimates are unlikely to have either the values nor the meanings which are needed. 

The coefficient estimates result from finding a set of values that can make a specified code model produce a simulated sample more like an observed sample than are samples from the other specified code models which have been tried. There is no guarantee that either sample used in a comparison will be representative of the ensemble from which the model samples or of the experimental procedure. No algorithm of any kind can sort out which features of any one random sample are characteristic of the ensemble, and which are idiosyncratic to that particular sample. Risks of non-representativeness can be reduced by using larger samples of experimental observations. As the number of simplifying statistics $p$ increases, the $n$ required to maintain the same risk of a non-representative sample increases exponentially. 

Even if the two samples are representative of the two ensembles which have been sampled, an ABC fitting algorithm may not converge to a useful estimate. ABC algorithms are based on rejection sampling, which is controlled by a tolerance parameter. If the tolerance parameter is set low enough, most candidate sets are rejected for inclusion in the posterior sample. If too many are rejected, there may be few or no observations in the posterior sample by the time that simulation resources are exhausted. If the tolerance is raised too high, the resulting sample is from a mixture of the prior distribution and the posterior. In the worst case, setting the tolerance too high produces a sample from the prior distribution on the coefficients which was chosen by the researcher and is not based on the experimental data at all. 

Even if the ABC algorithm converges to a useful estimate, the estimate may only provide a good model for the joint distribution of the fitting values. While an ABC algorithm may be able to fit a \textit{general} code model using only the information from $p$ fitting variables, this can only happen if those $p$ fitting variables are sufficient for the coefficients being estimated. It is impossible to prove sufficiency for any set of simplifying statistics, since it is impossible to write down a simple analytic expression for the likelihood of a code model. 

The construction of a general RT code model consists of using separate model components to represent disorder, unpredictability, and uncertainty. These components cannot be fit separately from each other. If coefficients from two or more of these model components are fit by an algorithm that compares samples of experimentally observed and simulated data, there is no algorithm that can identify which aspects of the two samples should be used to fit the coefficients from the disorder, unpredictability, or uncertainty models. Any such algorithm will make use of whatever capacities the model has to represent the observed sample, regardless of what the modeller intended the model components to represent. Even if a robust and highly useful model arises from the fitting process, it is likely that none of the coefficient estimates will be well-defined and interpretable estimates of the quantities that they were intended to represent. 

If the general RT code model cannot be shown to be elementally representative, the most accurate description of the coefficient estimates is that they specify a restricted RT code model that fits better than one based on an initial estimates of the coefficient values. The coefficient estimates specify the model that best performs out of the few that were tried, but those estimates may be far from the coefficient values which would provide the best agreement between the samples via the chosen metric. When estimates are produced by an algorithm, the algorithm 
may make the best use of available resources to make the fit, and will minimize the biases that can arise from excessive preference for expert judgment over interaction with observations of the phenomenon. The fitting procedure alone cannot establish the validity of model outputs, and can only produce fitted restricted RT code models which may be worth assessing for validity. 

\section*{Validating a fitted RT code model}

The process of validating a fitted restricted or general RT code model is dependent on the use to which the model will be put. Validation could range from only a careful analysis of how the model was built to an extensive program of repeated experiments involving many person-years of work after a functional general RT code model is initially contributed. What is needed to validate will depend on which known weaknesses of the model need to be investigated. It will also be necessary to ensure that any validation methods build up evidence of validity from experimental data, and not just by looking for a lack of disagreement with current theoretical narratives and historically useful models. 

\subsection*{Speculative models}

In some circumstances, it is useful to speculate about what might emerge if some aspects of elements of a phenomenon were constrained to act in some particular simple and plausible way. These models suggest what might happen, and have value when their outputs contradict explanatory narratives and expert judgments about how realizations of a phenomenon ought to emerge from their constitutive elements. There is no need to compare model output to data in order to answer a what-if question. The primary utility of a speculative model lies in motivating new experimental investigations of the phenomenon, and in exposing weaknesses of existing theory. 

A classic example of a speculative model is the original SIR model for an epidemic \cite{anderson:1991}.  The assumptions about how the modelled disease would progress through a human population are not true of any disease or of any human population. When these assumptions were made, they allowed the equations of simple compartment model to be interpretable as representing some aspects of an epidemic spreading through a human population. Those equations had a simple formal solution that could be easily found and understood. The analysis of that solution suggested that if an epidemic of that kind were possible, then its spread could be prevented by immunizing a critical proportion of the population and leaving the remainder non-immunized. This counteracted the belief that complete immunization was necessary to prevent an epidemic, and motivated public health investigation herd immunity. 

\subsection*{Explanatory and Emulative Models}

If a speculative model can produce output, then it may be possible to validate that model as one which produces strong evidence in support of an explanation for how realizations of the phenomenon emerge from elemental interactions. It is more likely that it will only be possible to find evidence that the model can emulate the distributions of some unpredictable aspects of realizations of the phenomenon. An explanatory claim may be that a PDE model for a pattern-forming system describes every important aspect of how individual disordered patterns are generated from interactions of the chosen elements, or that a particular simple mechanism is almost always responsible for the emergence of some important aspect of a powder flow. In order to provide this kind of evidence, the fitted restricted RT code model must be able to emulate the joint distribution of all important observable statistics that describe the aspects of the phenomenon which need to be explained. Since there is a risk that a fitted RT code model could be a good emulator while being elementally false, the most important part of explanatory validation is to provide good evidence that the model is elementally representative.

If all modelled attributes of elements can be accurately and precisely observed in an experimental realization as they interact over time, it may be possible to use experimental observations to argue that a verified general RT code model is elementally representative. Since these element-scale data are almost never observable, fitted RT code models intended to support an explanation need to be validated by other means.

\subsubsection*{\textit{A priori} validation of explanatory models}

In order to obtain the resources to construct and fit a restricted RT code model, it is necessary for its constructors to justify that they are using scarce resources to build a model which may prove to be useful. During the construction, choices of what to include of exclude will be grounded by theoretical narratives which seek to explain what has worked in similar models for similar phenomena. Trusted numerical routines for solving linear and non-linear equations will be used. If some parts of the model are innovative, then there may be an empirical or theoretical argument that justifies the innovation. The innovation may also be justified based on previous successes of the researchers who proposed it. When funding and resources to build an explanatory model are sought, all of this can be presented as a narrative explaining why the model might have the capacity to be elementally representative. Many of these justifications are constructed before any work is done to construct the code, to obtain any simulated observations, or to assess the quality of those simulated observations. If this kind of justification of potential validity is used as the sole reason for claiming that an RT code model is elementally representative, this is \textit{a priori validation} that the RT code model is elementally representative. 

\textit{A priori} validation of elemental representativity may be stated as an explicit statement of faith that a new RT code model must be elementally representative, based on expert judgment of the utility of similar methods for other problems and based on the reputations of the researchers who constructed it. It may also be undertaken implicitly, by treating a general RT code model as if it were elementally representative as soon as it can be made to produce output. If the initial outputs of a fitted RT code model are considered to be strong evidence of the validity or invalidity of the theoretical narrative that inspired the model's construction, then that fitted RT code model has been implicitly \textit{a priori} validated. 

\textit{A priori} validation has greatly simplifies model validation and fitting. Since the coefficients in the RT code model and the formal model can be taken to have the same meanings, values of many of the coefficients can be fit before gathering any data through use of values that were found from useful models of other experimental phenomena. The number of coefficients to be fit is minimized, and in some cases there may be no need to fit any coefficients using data. It may also be taken to be true that the fitted coefficient values should produce good emulation for any small set of simplifying statistics that are found from the fitted general RT code model output. In these circumstances, any visualizations that look right and any successful emulations of samples of small collections of simplifying variables can be considered to be confirmatory evidence of fitted general RT code model validity. If there is evidence that the fitted model fails to emulate the distribution of some collections of simplifying variables, this is evidence against the proposed explanation. If tweaking the values of some coefficient estimates greatly improves emulation performance, then this also is evidence against the proposed explanation. 

In the absence of being able to find evidence that the model is elementally representative though accurate and precise observation of all of the interacting elements in a realization of an experimental phenomenon, \textit{a priori} is validation by assumption. If the RT code model is in fact elementally false, \textit{a priori} validation could lead to a potentially useful emulator being rejected if tweaking the coefficients greatly improves the quality of emulation. \textit{A priori} validation may result in a proliferation of many models based on different mechanisms which are elementally false but which are all treated as being strong evidence for the explanations that they are believed to support. 

\subsubsection*{Data-centric validation}

If direct observation of the elements in an experimental realization is impossible, then an alternative to \textit{a priori} validation is to assume that a fitted RT code model is elementally false and then to accumulate evidence in favour of it being elementally representative by examining its performance. In every step of this process, it would be necessary to avoid any shortcuts that would avoid critical investigation of serious weaknesses in the fitted RT code model. 

If a fitted general RT model were elementally representative, then it would be expected to have certain properties. It would provide random  samples from the distributions of the simplifying statistics that were used to fit it via a restricted RT code model. If fitted general RT code model output were used to produce simulated samples of simplifying statistics that were not used in the fitting process, then it would be expected that many of these samples would have statistically indistinguishable distributions from experimental observations of those statistics.
When any small combination of simplifying statistics are chosen, the model should be able to emulate their joint distribution. Similar utility of emulation for those joint distributions should also be found from a general RT code model that has been fit to observations of similar phenomena thought to emerge from the same kinds of elemental interactions. 

The first step in a data-centric validation would involve finding no strong evidence that the fitted restricted RT code model fails to emulate the distribution of its fitting statistics for a single experimental phenomenon. The double negative in the preceding statement is necessary, since the only evidence for emulation comes from random samples. If no difficulties are initially found, the next step would be to see how far that capacity to emulate extends out to new observations of the same phenomenon. If the coefficients of a fitted restricted RT code model were based on a small set of simplifying statistics, it would be necessary to show that the joint distributions of some other observable simplifying statistics for the phenomenon can also be emulated. If the fitting procedure is applied to a different phenomenon (say, the original experimental sphere packing process using spheres made from a slightly different material or spheres that were slightly smaller), good emulation of joint distributions for the original fitting statistics and others would also be useful evidence. If good emulations for several phenomena allowed the development of models which could emulate the joint distributions of small collections of those statistics for as-yet-unobserved experimental phenomena, this would provide stronger evidence. If the model and its fitting procedure could be robustly and reliably extended to some collection of well-defined repeatably observable phenomena which occur outside of experimental settings, that would be even better evidence. If a general RT model and its fitting procedure had all of these properties, then that great demonstration of utility would provide no evidence against the general RT code model being elementally representative for the phenomena whose simplifying statistics it can emulate. 

If \textit{a priori} validation is used, this kind of utility of the general RT code model could be deduced, and not seeing it would be evidence of model failure. If this kind of utility is observed in data-centric validation, then taking this utility to be evidence of the general RT code model as being elementally representative for some class of phenomena is justification by abduction, not deduction. If the first paper written about a fitted RT code model shows that the model can emulate some features of a single random sample of observations, this is an absence of any  evidence from the data that the proposed explanation is wrong, rather than strong confirmatory evidence of the proposed explanation. When evidence accumulates for the broad utility of the fitted model as an emulator, it is evidence that many different collections of simplifying variables have model-based joint distributions which cannot be distinguished from experimental joint distributions by means of statistical inference. This does not exclude the possibility that the model may yet fail for new and important simplifying statistics which have not yet been observed, nor does it exclude the possibility that the models for disorder, uncertainty, and unpredictability are each failing to represent those aspects of the experimental phenomenon that they are intended to represent. It may never be possible to validate a model as being elementally representative using data-centric approaches based on successful emulation of joint distributions of simplifying statistics, but making the effort to do may result in an RT code model which is so broadly useful that it can be treated as being elementally representative for some simplifying statistics with no significant cost to a researcher in the short term.

Working towards a data-centric validation will require immense amounts of experimental investigation that will be more than what a single research group can achieve or document within a single publication. Instead, the value will need to emerge through persistent demonstrations of utility described in reviews of reports from many different researchers, long after the first working example of a restricted RT code model is  first presented in a publication. Demonstrations of utility require data from new experimental realizations which cannot be replaced by simulated data, since the use of simulations to substitute for experiments would require \textit{a priori} validation of the models that produce the simulated data. Since no single research group will be able to carry out a complete program, the ability to publicly share simulated data sets and reproducible general RT model code is essential. 

In data-centric validation of an explanation, the validation process must provide evidence against the initial assumption that the fitted restricted and general RT code models may be elementally false but still useful as emulators in a limited range of circumstances. This requires actively seeking out evidence of how a fitted RT code model fails at emulating the distributions of other simplifying statistics that can be observed on the realizations of the phenomenon. If important aspects of the experimental phenomenon can only be partially described by a collection of simplifying statistics, it will be necessary to develop new simplifying statistics and new means of observing them in experimental realizations. Looking for strong evidence of model failure in this context is as important as initial findings that the model might be able emulate something. This must be recognized in the context of publication, researcher evaluation, and research funding. 

\subsection*{Validating emulation}

Unless a strong case can be made that a general RT code model is elementally representative, it must be considered to be elementally false. The models may still be useful if it is possible to provide evidence that they can sample from multivariate distributions of simplifying statistics which are not distinguishable from the distribution of those statistics for the experimental phenomenon by means of statistical inference.  Unlike validation of explanatory models, validation of an emulator can be established when enough evidence has accumulated regarding the utility of its simulations.

It is first necessary to establish that the fitted restricted RT model can emulate the distribution of the simplifying statistics used to fit it. This distribution will be multivariate and should not be expected to be Gaussian. It should also be expected that the statistics are not independent of each other, and so the simulation should at least be able to produce samples with the similar values for measures of typical behaviour, of variability, and of pairwise variable dependence to those found from experimental observations. It may be possible to summarize the joint distribution using a multivariate Gaussian distribution or a copula-based model and to find no evidence  that samples from the model and the experiments  are statistically distinguishable when summarized in this way. 

It is necessary to investigate the degree to which the fitted restricted RT code model can be extended to emulate the distributions of other simplifying statistics not used in the fitting process. This will be useful when more statistics need to be simulated than are needed to fit the model, or when an experimentally as-yet unobservable simplifying statistic needs to have its distribution investigated. In the latter case, the fitting variables and perhaps several others should be chosen from simplifying statistics which are expected to be dependent on the unobservable statistic. The failure to extend emulation to the distributions of other variables will provide evidence of the limitations of the general RT code model.

It is necessary to consider how well the emulation can extend to other very similar experimental phenomena. In the case of sphere packings, it may be possible to change the sphere size of the material out of which the spheres are made. If a new RT code model is fit after small changes to the experimental phenomenon, then a failure to emulate previously well-emulated distributions of simplifying statistics will provide strong evidence against the broad utility of the general RT code model.

\subsubsection*{Validating experimental reproducibility}

Before validating a fitted restricted RT code model, it is necessary to first validate the replicability of the experimental phenomenon. Since the phenomenon is complex, it is likely that its realizations are affected by variables which are neither observed nor controlled. An experimental procedure needs to be simple enough to implemented, but may not be complete enough to ensure that the same phenomenon is being created and observed by the same researchers at a later time, or when using different apparatus, or when the procedure is followed by a different group of researchers in a different location. 

If an attempt to replicate a series of experimental realizations is undertaken, the new random sample needs to be compared to the initial one and to any other samples from previous replications. It will be possible to compare pairs of samples using statistical tests which can seek evidence of differences in some aspects of the multivariate distributions of the fitting statistics. When these tests are used on small samples, there will be significant risks that the tests will either fail to detect differences or find evidence of differences which do not exist. As a result, experiments will need to be repeated by many different research groups at different times in order to amass enough observations to make a strong case for experimental replicability. If such a case can be made, it may be acceptable to combine some or all of the samples into a larger one which is less likely to be unrepresentative than any of its constituents. If samples are pooled without looking for evidence of that contradicts the assumption that the experiment is indefinitely replicable if the procedure is followed, then a combined sample may be a mixed sample from many fundamentally distinct experimental phenomena. Under those circumstances, the mixing would not be replicable and so no clear conclusions about any one of the mixed phenomena could be obtained. 

If no evidence is found of differences in samples of observations of fitting variables from distinct experiments, then this does not imply that other simplifying statistics will have the same distribution for different experiments. If there are other simplifying statistics which are to be emulated via the fitted general RT code model, it would be necessary to look for evidence of differences in distributions for these statistics between experiments. If there are other observable simplifying statistics which describe aspects of interest and which are correlated with the fitting statistics, these would also be useful to investigate. 

\subsubsection*{Validating emulation of the joint distribution of fitting statistics}

If no strong evidence can be found that the experiment is not replicable with respect to the fitting variables, then the model's capacity for emulation can be evaluated using a similar inferential approach. The initial sample of simulated observations can only be compared to the experimental observations used to fit the model. This initial sample would be expected to have the best agreement with the model versus all other samples on account of overfitting. Strong evidence of differences between simulated and experimental distributions at this stage would be troubling, but if samples were small those differences might be explained by sampling error.  Validation requires new samples of realizations of the experimental phenomenon which were not used to fit the model. These new samples can be fit separately and the two sets of coefficient estimates can be compared. The new sample can be compared to simulated data from the initial fitted model, and vice versa. As new samples of observations are obtained from new experimental realizations, there also needs to be a well-thought-out procedure for combining the information from these samples and the models fit from them. This procedure also needs to accommodate the difficulties which arise when the same samples are used in many different statistical tests. 

For complex phenomena and their RT code models, validation of emulation is a long-term process rather than a single quick and decisive computational act of confirmation. Each contribution based on a single replication of the experiment can provide some evidence for or against validity, but no single experimental fit can confirm validity of a fitted restricted RT code emulator. Research communities that need validation need to develop methodology that can make an efficient community-based validation process possible. All experimental observations and simulated samples need to be freely available. The code model must be available in a portable form whose reproducibility can be confirmed, and it needs to be accompanied by clear documentation that can prevent avoidable problems with reproducibility. When statistical tests are undertaken, the testing procedures need to be documented and the results made publicly available regardless of whether or not the results support validity. It may also be useful to pre-register experiments, so as discourage the concealment of results which  contradict the expectations of the researchers who produced them. 

The validation process has no clear endpoint. While a single sample and its fitted model might provide extremely weak evidence of emulator validity, there are resource-based constraints on how many times an experiment can be repeated. There is no guarantee that what can be done will lead to the accumulation of enough evidence to make a useful assessment of validity. 

It must also be understood that the failure of a particular restricted RT code model to be a valid emulator of an experimental phenomenon is as important as finding no evidence against validity. Having a record of what failed is useful if researchers decided to fit the general RT code model using different fitting statistics, or if they seek to use new fitting methods. The general RT model code may prove to be useful for modelling other phenomena, even if it proved to be useless in its initial intended applications. 

\subsection*{Extension of emulative validity to other phenomena}

If a restricted RT code model can be validated for its fitting statistics for one replicable experimental phenomenon, it may be possible to fit that same model to other similar phenomena. Since an elementally false model is a mathematical device rather than a representation of the emergence of the distributions of simplifying statistics from elemental interactions within realizations of the experimental phenomenon, it cannot be assumed that validity can be extended to new phenomena. 

For each new phenomenon, all of the steps in validation for the original phenomenon need to be repeated. There need to be multiple replications of experiments to provide evidence for or against experimental reproducibility. There needs to be coordinated and systematic comparison of how the fitted models perform on data not used to fit them. This process may consume less resources than the initial validation, since knowledge gained from that initial validation process may make the validation of the model for the new phenomenon more efficient. 

\subsection*{Extension of emulative validity to other variables}

If a general RT code model is fit using coefficients from a restricted RT code model, then that general RT code model may be used to simulate samples of other simplifying statistics that were not used in the fitting process. If those statistics are not observable in the experimental phenomenon, anything concluded from simulations of those variables needs to be considered to be speculative, regardless of how well the fitted restricted RT model can emulate the joint distribution of the fitting statistics. If the new statistics are observable, then the model's ability to emulate their distributions needs to be validated in the same way as the fitting statistics were validated.

It is first necessary to validate that a new simplifying statistic has the same distribution in further repetitions of the experiment. If the experimental procedure is sound enough to make observations of the fitting statistics reproducible, that reproducibility may not extend to any other observable simplifying statistics. Once there is evidence of experimental reproducibility, multivariate tests for evidence of differences in joint distributions can be used to compare the simulated and observed samples of the original fitting variables together with the new statistics. 

Since success in emulation for some useful simplifying statistics may encourage use of the general RT code model in many other settings, there is great value in collecting and then publicly releasing more information than is needed to validate a particular model. This information could consist of experimental measurements, images, movies, or other data from remote sensing, together with the procedures for constructing them. This information could provide observations of other variables to be used in the fitting and validation of other RT code models. 

\subsection*{Establishing the limits of emulation}

Since the general RT code models are initially assumed to be elementally false, there are no grounds for believing that their capacity to emulate extends indefinitely beyond the collections of experimental phenomena and of simplifying statistics for which validation of emulation has been successful. To show broad utility, there needs to be an ongoing effort to investigate previously unobserved simplifying statistics for as-yet-unexamined phenomena. 

If the general RT code model is elementally false, there will be some simplifying statistics whose distributions cannot be emulated. This may arise from problems with the ability to usefully model measuring and imaging instrumentation, but also may result from the modelled elements not having the capacity to generate an emergent distribution for the simulated simplifying statistics that resembles the distribution of experimental observations. Public knowledge of which distributions the general RT code model cannot emulate is as valuable as evidence of what the model can emulate. Failures of the model to emulate the distributions of particular simplifying statistics may also suggest how the general RT code model can be improved. 

Efficient investigation of the limitations on emulative capacity for fitted RT code models requires the availability of a large collection of many diverse simplifying statistics. The collection should consist of statistics which are at best weakly dependent and which summarize many diverse aspects of the general RT code model output beyond those of interest at the time of fitting. As many as possible should be observable on experimental realizations of the phenomenon. 

If specific model weaknesses are suspected then statistics can be sought or developed which may be able to quantify or categorize those weaknesses. If many different statistics can describe an important aspect of a phenomenon and only some of them are used in fitting, the others could be chosen as diagnostics for model limitations. If samples of experimental and simulated observations of many different simplifying statistics are available from a fitted general RT code model, then efforts can be made to find out which of these statistics could be used to create a classification rule that would be highly successful at classifying each observation as being experimental or simulated. If such a classifier could be built, then its success rate and the statistics selected to build it would suggest which simplifying statistics have distributions which are not being properly emulated. If no such classifier can be built, this may be because the simplifying statistics needed to identify the failings of the model have not yet been developed. 

\subsection*{Comparing different fitted RT code models}

The statistical methods used to seek evidence of problems with reproducibility or emulation can also be applied to compare samples from two different specified RT code models. If at least one of these models has not been validated as a good simulator of observations from the experimental phenomenon, then any such comparisons will be speculative in terms of what they imply about the experimental phenomenon but informative about differences between specified RT code model outputs.

Comparing samples from different specified RT code models is easier than the other two forms of comparison since there is no requirement that a simplifying statistic be observable on an experimental realization. Comparison studies between models may suggest where efforts should be made to develop both new simplifying statistics and the methods needed to observe their values in experimental realizations. If simulations are fast and can be produced in greater numbers than experimental realizations, then many of the risks associated with model evaluation based on small random samples can be reduced. 

Comparisons between specified code model outputs are useful when trying to anticipate what kind of impact may result from a change to a model or a statistic. If a simplification to a model is proposed in order to make the code model run faster, the distributions of fitting statistics can be compared to seek evidence of differences in distributions. It is also possible to compare the resulting fitted general RT code model outputs by calculating many diverse weakly associated statistics and trying to build a supervised learning model that can distinguish between the outputs of the two models. In the sphere packing example, a new model might have an interaction mechanism simplified, or eliminate all of the terms related to angular momentum. Having the good knowledge of how such changes affect model output is necessary to learn about  the non-obvious costs of making the code model run faster. 

Specified RT code model output may also be useful for developing and evaluating new simplifying statistics. A new statistic can be imagined and then formally defined based on trying to quantify some aspect of the model that theory says could be important, or based on trying to quantify something seen in an image. For some choices of simulated disordered patterns, it may be possible to obtain evidence that a new statistic can identify differences between samples of simulated patterns that other statistics cannot. It may also be able to provide evidence that a simple-to-calculate statistic is as useful as one which requires more effort to compute. 

\subsection*{Images in diagnostics}

All discussion of validation so far has focused on simplifying statistics which provide some form of numerical value or categorical label for each realization. A collection of these statistics map the very-high-dimensional general RT code model output down to a sufficiently small  collection of values and labels that can be used in fitting and validation procedures. The calculation of these statistics is a dimension reduction process which discards a very large amount of information. The production of still and moving images from fitted RT code model output is a less destructive form of data reduction, but one whose output is useful in different ways from the utility of simplifying statistics.

Visualizations produced from general RT code model output are idealized versions of what is seen when a realization of a phenomenon is seen. They provide a way to compare simulated and experimental observations that can show much more complex similarities and differences than small sets of numbers can.  

The process of looking for similarities and differences between images using human vision is a process which results in further loss of information. The human visual process is not a camera, but one in which the mind consciously and unconsciously interacts with neural responses of retinal cells \cite{frisby}. Not every difference between two images is visible, and the human mind can perceive illusory details of seen images which are not present in the image itself. If no differences between two images are seen, there may still be information in those images which can retrieved through use of statistics which simplify image data. This can provide further incentive to develop more simplifying statistics that can extract these invisible features from complex patterns. 

When fitted general RT code models are elementally false, the interactions of the elements as modelled may be profoundly different from how the elements interact in an experimental realization. A general RT code model may produce visualizations which look wrong, even when the fitted RT code model is successful at emulating the joint distribution of some simplifying statistics. This kind of difference must be accepted as a possible outcome from using an elementally false model. If improvements to an elementally false RT code model diminish the visible problems, this might suggest progress towards the development of an elementally representative model. 

\section*{Scientific use of RT models}

Random trajectory code models provide a way of modelling some aspects of complex phenomena when no simple models are available. They can simulate samples of aspects of individual realizations in a way that may be able to capture both the complex internal structure of each realization and the ways in which individual realizations differ from each other. By using simple models on one scale, they provide a way to simulate emergent complexity on a larger scale. They are needed when methods from statistical physics are incapable of usefully representing important aspects of the phenomenon of interest. 

RT code models are models which are code, not formal mathematics. Models are fit and validated by comparing simulated realizations generated by code with observations of experimental realizations. The RT code model is not necessarily an approximation to the original system of differential equations which inspired it. When elemental representativity cannot generally be directly validated, the general RT code model is essentially a non-linear probabilistic black-box model in code which may have the capacity to usefully emulate the distributions of some unpredictable aspects of the phenomenon if a useful set of coefficients exist and can somehow be found. 

The RT code model development pathway shows that there are many unverifiable assumptions which, if false, would make the general RT code model elementally false. The coefficients of the general RT code model would not necessarily have the meaning that they possessed in the original formal model, and would need to be fit based on choosing values which make simulated realizations as similar as possible to observed ones. This similarity would be assessed based on a choice of a limited number of summarizing statistics. For an elementally false model, there would be no guarantee that coefficient estimates would not greatly change if the summarizing statistics used to fit the model were changed. Fitting would also require simplifying statistics which can be both observed and simulated, which may not be the statistics that are needed to fit a model that is useful. If small samples must be used in the fitting process, then there is a risk that those samples will be unrepresentative of the specific model or of the phenomenon and there will be limitations on how many coefficients can be fit using experimental observations. This will require the use of expert judgment to choose whatever coefficient values cannot be fit using experimental data. When a new fitted model is presented, there is a very real possibility that it will prove to be less useful than later fits based on new observations and different strategies for fitting. The existence of a fitted, output-producing model cannot alone suggest that the output will be able to usefully simulate experimental outcomes.

Validation of a fitted RT code model as emulative for the joint distribution of some collection of simplifying statistics requires many comparisons between samples from the fitted model and new samples of experimental observations. Most of this needs to follow the initial contribution of the general RT model code and an example of a fitted model to the research community. Establishing emulative capacity  requires review of subsequent attempts at improved fittings of the model and the combination of experimental estimates of coefficients by meta-analysis. All of this requires coordinated efforts between research groups for greatest efficiency. If it is possible to establish evidence that a fitted general RT model can predict many simplifying statistics well, that model may have some kind of value in justifying an explanation for how the behaviour of the simplifying statistics emerge from elemental interactions. If there is little to no data-based evidence that a fitted RT code model can predict new observations of some simplifying statistic, then it is only useful for summarizing the data to which it was fit. Under these circumstances, the model user must acknowledge that there may be no natural phenomenon that the model can represent.

All of the preceding comments implied that model users were incorporating probabilistic models for uncertainty or unpredictability which could be used to provide random samples from the model. If a purely deterministic code model is built for the disorder alone and run for only one set of initial conditions to produce a single simulated realization, then this model is next to useless except as a demonstration that the code runs. A single exception to this would be in cases where the goal of modelling was to study some aspect of the phenomenon which was essentially invariant between realizations. If some collection of simplifying statistics could be found that was invariant between realizations and observable in both experimental and simulated realizations, then it might be possible to fit a reduced RT code model by finding a set of coefficient values that made the simulated values of the invariants equal to the observed values. Any pure disorder code model fit in that way would be unable to represent any feature of the phenomenon which did vary between realizations. While values of non-invariant quantities could be found or observed in a single realization, this would be sample of size one of a random variable and as such would provide essentially no useful information about the distribution of that random variable. Since many important aspects of complex phenomena do vary between realizations, use of invariants for fitting would be unwise, although some might be useful for model validation. Invariants could be used if there was strong evidence that the general RT code model was elementally representative. 

If the general RT code model was elementally representative, most of the problems previously noted would disappear. Fitting could be done using summarizing statistics which are invariant between realizations, and the model could be trusted to have the capacity to emulate the distributions of many more simplifying statistics other than the ones used to fit it. The facts that an elementally representative model is much easier to fit and validate, that RT code models are usually the only models available, and that validated RT code models are needed do not combine to provide evidence that an RT code model is elementally representative. \textit{A priori} validation of fitted RT code models amounts to assuming that the model is elementally representative based on little to no empirical evidence, or to simply ignoring all ways in which the process of constructing a general RT code model is very likely to produce an elementally false model that may have no capacity to emulate or explain any aspect of the phenomenon. If a research community acts as if its general RT code models are elementally representative without evidence to back that claim, then the value of fitted RT code models as providers of evidence in favour of explanations may be grossly overstated. When that happens, model use may hinder rather than facilitate the development of better understanding of the phenomenon. If a research community privileges assuming elemental representativity over assessment methods which use data to actively seek evidence of model weaknesses, it is reasonable to ask if the use of RT code models represents scientific knowledge-building or a retreat to more empowering pre-scientific methods knowledge-building. 

This discussion of RT code modelling has been strongly focused on the weaknesses of these models. While acknowledgement of these weaknesses may result in the more efficient use of these models by research communities in knowledge-building, these weaknesses have also be exploited by those who find scientific knowledge a threat to their ability to market tobacco, deplete aquifers, or continue mining coal \cite{oreskes:2010}. Understanding how to develop the most effective scientific methodology for using RT code models requires direct engagement with aspects of knowledge-building that are not accessible by mathematical or computational analysis, and which can only be partially confronted by within-community criticism of research community methodology. It will require recognizing that useful scientific knowledge is constructed by the collective activity of human researchers working to meet human needs. Understanding how research communities achieve this and using that knowledge to develop an effective methodology for using RT code models will require engagement with the sociology, history, and philosophy of science \cite{ideasB}.

\bibliography{/Users/jpicka/Desktop/spc20.bib}

\end{document}